\documentclass{article}

\usepackage[nonatbib,preprint]{neurips_2023_modified}
\usepackage[numbers,sort&compress]{natbib}

\usepackage[utf8]{inputenc} 
\usepackage[T1]{fontenc}    
\usepackage{url}            
\usepackage{booktabs}       
\usepackage{amsfonts}       
\usepackage{nicefrac}       
\usepackage{microtype}      

\usepackage{caption}
\usepackage{subcaption}
\usepackage{wrapfig}
\usepackage{etoolbox}
\usepackage{verbatim}
\usepackage[inline]{enumitem}
\usepackage{amsmath}
\usepackage{graphicx}
\usepackage{tabularray} 
\usepackage{caption}
\usepackage{multirow}
\usepackage{makecell}
\usepackage{bbm}

\usepackage{colortbl}

\usepackage[usenames,dvipsnames]{xcolor}
\usepackage{hyperref}
\hypersetup{ %
	pdfauthor={},
	pdftitle={},
	pdfsubject={},
    pdflang={en},
	bookmarks,
	bookmarksnumbered,
	backref=page,
	pageanchor=true,
	plainpages=false,
	colorlinks=true,%
	linktocpage=true,
	citecolor=MidnightBlue,%
	filecolor=BrickRed,%
	linkcolor=NavyBlue,%
	urlcolor=NavyBlue 
}

\usepackage{tikz}
\usetikzlibrary{arrows,positioning,shapes}
\tikzset{box/.style={rectangle, draw=black, minimum size=0.25cm}}

\usepackage{xspace}
\usepackage[title]{appendix}

\newcommand{\textcode}[1]{{\fontfamily{cmtt}\selectfont #1}\xspace}

\newcommand{\TechChatGPT}{\textcode{ChatGPT}}
\newcommand{\TechGPTFour}{\textcode{GPT-4}}
\newcommand{\TechTutor}{\textcode{Tutor}}
\newcommand{\Evaluators}{\textcode{evals}}

\newcommand{\PGCD}{\textsc{GCD}}
\newcommand{\PFibonacci}{\textsc{Fibonacci}}
\newcommand{\PDivisors}{\textsc{DivisorsDiv3}}
\newcommand{\PPalindrome}{\textsc{Palindrome}}
\newcommand{\PMergeStrings}{\textsc{MergeStrs}}

\newcommand{\BPaGCD}{\textnormal{BP-01}}
\newcommand{\BPbGCD}{\textnormal{BP-02}}
\newcommand{\BPcGCD}{\textnormal{BP-03}}
\newcommand{\BPdGCD}{\textnormal{BP-04}}
\newcommand{\BPeGCD}{\textnormal{BP-05}}

\newcommand{\BPaFibonacci}{\textnormal{BP-06}}
\newcommand{\BPbFibonacci}{\textnormal{BP-07}}
\newcommand{\BPcFibonacci}{\textnormal{BP-08}}
\newcommand{\BPdFibonacci}{\textnormal{BP-09}}
\newcommand{\BPeFibonacci}{\textnormal{BP-10}}

\newcommand{\BPaDivisors}{\textnormal{BP-11}}
\newcommand{\BPbDivisors}{\textnormal{BP-12}}
\newcommand{\BPcDivisors}{\textnormal{BP-13}}
\newcommand{\BPdDivisors}{\textnormal{BP-14}}
\newcommand{\BPeDivisors}{\textnormal{BP-15}}

\newcommand{\BPaPalindrome}{\textnormal{BP-16}}
\newcommand{\BPbPalindrome}{\textnormal{BP-17}}
\newcommand{\BPcPalindrome}{\textnormal{BP-18}}
\newcommand{\BPdPalindrome}{\textnormal{BP-19}}
\newcommand{\BPePalindrome}{\textnormal{BP-20}}

\newcommand{\BPaMergeStrings}{\textnormal{BP-21}}
\newcommand{\BPbMergeStrings}{\textnormal{BP-22}}
\newcommand{\BPcMergeStrings}{\textnormal{BP-23}}
\newcommand{\BPdMergeStrings}{\textnormal{BP-24}}
\newcommand{\BPeMergeStrings}{\textnormal{BP-25}}

\definecolor{ExpHighlight}{rgb}{0.8, 0.1, 0.1}
\definecolor{mygreen}{rgb}{0,0.6,0}
\definecolor{mygray}{rgb}{0.5,0.5,0.5}
\definecolor{mymauve}{rgb}{0.58,0,0.82}
\definecolor{CodeHighlight}{rgb}{1,1,0.7}
\definecolor{CodeGray}{rgb}{0.8,0.8,0.8}

\definecolor{promptinputcolor}{rgb}{0.58,0,0.82}
\newcommand{\promptheader}[1]{{\large{\textcolor{RoyalPurple}{\textbf{#1}}}}}
\newcommand{\promptinput}[1]{{\textcolor{promptinputcolor}{\textcode{#1}}}}

\newcommand{\componentheader}[1]{{{\textcolor{RoyalPurple}{\textbf{#1}}}}}

\newcommand{\tutorcellcolor}{green!25}
\usepackage{listings}
\makeatletter
\let\old@lstKV@SwitchCases\lstKV@SwitchCases
\def\lstKV@SwitchCases#1#2#3{}
\makeatother
\usepackage{lstlinebgrd}
\makeatletter
\let\lstKV@SwitchCases\old@lstKV@SwitchCases

\lst@Key{numbers}{none}{%
    \def\lst@PlaceNumber{\lst@linebgrd}%
    \lstKV@SwitchCases{#1}%
    {none:\\%
     left:\def\lst@PlaceNumber{\llap{\normalfont
                \lst@numberstyle{\thelstnumber}\kern\lst@numbersep}\lst@linebgrd}\\%
     right:\def\lst@PlaceNumber{\rlap{\normalfont
                \kern\linewidth \kern\lst@numbersep
                \lst@numberstyle{\thelstnumber}}\lst@linebgrd}%
    }{\PackageError{Listings}{Numbers #1 unknown}\@ehc}}
\makeatother
\title{Generative AI for Programming Education: Benchmarking ChatGPT, GPT-4, and Human Tutors\thanks{This article is a full version of the poster (extended abstract) from ICER'23~\cite{icer23poster_ChatGPT_pythonprog}.}}

\author{
    Tung Phung\\
    MPI-SWS\\
    \texttt{mphung@mpi-sws.org}
    \And
    Victor-Alexandru P{\u a}durean\\
    MPI-SWS\\
    \texttt{vpadurea@mpi-sws.org}
    \And
    Jos{\'e} Cambronero\thanks{These authors are listed in alphabetical order. Correspondence to: Adish Singla <\texttt{adishs@mpi-sws.org}>.}\\
    Microsoft\\
    \texttt{jcambronero@microsoft.com}
    \And
    Sumit Gulwani\footnotemark[2]\\
    Microsoft\\
    \texttt{sumitg@microsoft.com}
    \And
    Tobias Kohn\footnotemark[2]\\
    TU Wien\\
    \texttt{tobias.kohn@tuwien.ac.at}
    \And
    Rupak Majumdar\footnotemark[2]\\
    MPI-SWS\\
    \texttt{rupak@mpi-sws.org}
    \And
    Adish Singla\footnotemark[2]\\
    MPI-SWS\\
    \texttt{adishs@mpi-sws.org}
    \And
    Gustavo Soares\footnotemark[2]\\
    Microsoft\\
    \texttt{gsoares@microsoft.com}    
}

\begin{document}

\maketitle

\begin{abstract}
Generative AI and large language models hold great promise in enhancing computing education by powering next-generation educational technologies for introductory programming. Recent works have studied these models for different scenarios relevant to programming education; however, these works are limited for several reasons, as they typically consider already outdated models or only specific scenario(s). Consequently, there is a lack of a systematic study that benchmarks state-of-the-art models for a comprehensive set of programming education scenarios. In our work, we systematically evaluate two models, ChatGPT (based on GPT-3.5) and GPT-4, and compare their performance with human tutors for a variety of scenarios. We evaluate using five introductory Python programming problems and real-world buggy programs from an online platform, and assess performance using expert-based annotations. Our results show that GPT-4 drastically outperforms ChatGPT (based on GPT-3.5) and comes close to human tutors' performance for several scenarios. These results also highlight settings where GPT-4 still struggles, providing exciting future directions on developing techniques to improve the performance of these models.
\end{abstract}

\section{Introduction}\label{sec.introduction}
Generative AI and large language models (LLMs) have the potential to power next-generation AI-driven educational technologies and drastically improve the landscape of computing education. We focus in this paper on the role of LLMs in enhancing introductory programming education. State-of-the-art models like OpenAI's ChatGPT~\cite{ChatGPT} and GPT-4~\cite{GPT4} could enhance programming education in various roles, e.g., by acting as a personalized digital tutor for a student, as a digital assistant for an educator, or as a digital peer for collaborative learning~\cite{DBLP:journals/corr/abs-2303-12712,baidoo2023education,LIM2023100790}. 
In our work, we seek to comprehensively evaluate and benchmark state-of-the-art LLMs for various scenarios in programming education.

Recent works have studied several LLMs for different
scenarios relevant to programming education~\cite{DBLP:conf/ace/Finnie-AnsleyDB22,zhang2022repairing,DBLP:conf/icer/SarsaDH022,leinonen23sigcse,edm23-pyfixv}. However, these works are limited for several reasons: they considered models that are already outdated (e.g., OpenAI's Codex~\cite{DBLP:journals/corr/abs-2107-03374} is no longer publicly available since March 2023), or they typically considered only specific scenario(s) (e.g., generating explanations). Consequently, there is a lack of systematic study that benchmarks state-of-the-art models for a comprehensive set of programming education scenarios.

In our work, we systematically evaluate two models, ChatGPT (based on GPT-3.5) and GPT-4, and compare their performance with human tutors for a variety of programming education scenarios. These scenarios are designed to capture distinct roles these models could play, namely digital tutors, assistants, and peers, as discussed above. More concretely, we consider the following six scenarios: (i) \emph{program repair}; (ii) \emph{hint generation}; (iii) \emph{grading feedback}; (iv) \emph{pair programming}; (v) \emph{contextualized explanation}; and (vi) \emph{task synthesis}.

\looseness-1We evaluate the performance of different methods (i.e., LLMs and human tutors) using expert-based annotations involving a mix of quantitative and qualitative assessments. We conduct our evaluation using five introductory Python programming problems with a diverse set of input/output specifications. For each of these problems, we consider real-world buggy programs based on publicly accessible submissions from 
\emph{geeksforgeeks.org} platform~\cite{geeksforgeeks}; these buggy programs are picked to capture different types of bugs for each problem. Our results show that GPT-4 drastically outperforms ChatGPT (based on GPT-3.5) and comes close to human tutors' performance for several scenarios. These results also highlight that GPT-4 still struggles with more challenging scenarios, e.g., grading feedback and task synthesis, where the performance of GPT-4 is quite low compared to that of human tutors.

The rest of this paper is organized as follows. Section~\ref{sec.problemsetup} provides an overview of our evaluation setup and introduces the data used for evaluation. Sections~\ref{sec.programrepair}--\ref{sec.taskcreation} provide results for the above-mentioned six scenarios. Section~\ref{sec.conclusion} discusses some limitations of our current work and directions for future work.


\section{Evaluation Setup}\label{sec.problemsetup}
This section provides an overview of our evaluation setup, including the programming education scenarios, Python problems along with buggy programs, and the overall process used for evaluation.

\paragraph{Programming education scenarios.} 
In our work, we consider the following six scenarios in programming education, capturing different roles that AI-based educational agents could play in the form of digital tutors, assistants, and peers:
\begin{enumerate}[label={(\roman*)},leftmargin=16pt,parsep=4pt]
    \item \emph{Program repair}, i.e., fixing a student's buggy program~\cite{yi2017feasibility,zhang2022repairing,edm23-pyfixv}.
    \item \emph{Hint generation}, i.e., providing hints to a student to help resolve current issues~\cite{leinonen23sigcse,edm23-pyfixv}.
    \item \emph{Grading feedback}, i.e., grading a student's buggy program w.r.t. a given rubric~\cite{DBLP:journals/corr/abs-2106-07340,DBLP:conf/aied/FunayamaSMMSI22}.
    \item \emph{Pair programming}, i.e., completing an incomplete/partial program written by a student~\cite{CopilotWeb,DBLP:journals/corr/abs-2210-14306,DBLP:conf/icse/Imai22,DBLP:journals/corr/abs-2306-05153}.
    \item \emph{Contextualized explanation}, i.e., explaining specific parts in the context of a given program~\cite{DBLP:conf/icer/SarsaDH022,DBLP:conf/onward/PotterMJO22}.
    \item \emph{Task synthesis}, i.e., generating new tasks that exercise specific types of concepts/bugs~\cite{DBLP:conf/nips/AhmedCEFGRS20,DBLP:conf/aied/GhoshTDS22,DBLP:conf/icer/SarsaDH022,neurtasksyn}.
\end{enumerate}

\paragraph{Introductory Python problems.} We conduct our evaluation using five introductory Python programming problems summarized in Figure~\ref{fig.problemsetup.problems}. These are typical programming problems, and variants of these problems appear in several online programming websites, courses, and textbooks~\cite{geeksforgeeks,codeforces,liang2013introduction,python-course-ocw-mit}. We picked these specific problems as they capture a diverse set of programming and algorithmic concepts required to solve and vary in their input-output specifications; moreover, solution programs for these problems are short and comprise up to $10$ lines of code.

\paragraph{Real-world buggy programs.} For these five problems, we consider buggy programs based on publicly accessible submissions from 
 \emph{geeksforgeeks.org} platform~\cite{geeksforgeeks}. These problems are available on \href{https://practice.geeksforgeeks.org/}{https://practice.geeksforgeeks.org/} at the following links: (a) \href{https://practice.geeksforgeeks.org/problems/gcd-of-two-numbers3459/1}{problems/gcd-of-two-numbers3459/1}; (b) \href{https://practice.geeksforgeeks.org/problems/fibonacci-to-n0811/1}{problems/fibonacci-to-n0811/1}; (c) \href{https://practice.geeksforgeeks.org/problems/number-of-divisors1631/1}{problems/number-of-divisors1631/1}; (d) \href{https://practice.geeksforgeeks.org/problems/palindrome-string0817/1}{problems/palindrome-string0817/1}; (e) \href{https://practice.geeksforgeeks.org/problems/merge-two-strings2736/1}{problems/merge-two-strings2736/1}.\footnote{We use detailed problem descriptions available at these links when designing prompts. Moreover, we use the problem specifications and automated test suites available at these links to check the correctness of a program.\label{footnote.problemdescription}} For each of these five problems, we picked five buggy programs, i.e., a total of $25$ buggy programs as summarized in Figure~\ref{fig.problemsetup.programs}. We picked these buggy programs to capture different types of bugs for each problem and ensure that these programs are associated with submissions by different users; moreover, these programs vary in size, ranging from $4$ to $31$ lines of code.\footnote{We note that the \emph{geeksforgeeks.org} platform doesn't provide URL links to specific submissions for a problem. For this reason, we have only provided URL links for five problems. The description of bug(s) and the number of lines provided in Figure~\ref{fig.problemsetup.programs} give useful insights into how one could curate similar data sets from different resources to conduct future studies.}

\begin{figure*}[t!]
\centering
    \scalebox{0.78}{
    \setlength\tabcolsep{2.0pt}
    \renewcommand{\arraystretch}{1.4}
    \begin{tabular}{p{3.5cm}||p{9cm}||p{3.5cm}p{3.5cm}}
        \toprule
        \multicolumn{1}{c||}{\textbf{Problem ID}} &
        \multicolumn{1}{c||}{\textbf{Short Text}} &
        \multicolumn{1}{c}{\textbf{Input}} &
        \multicolumn{1}{c}{\textbf{Output}} \\
        \midrule
        \multicolumn{1}{c||}{\PGCD} &
        \multicolumn{1}{l||}{Given two positive integers $n_1$ and $n_2$, find GCD of $n_1$ and $n_2$.} &
        \multicolumn{1}{c}{Two integers} &
        \multicolumn{1}{c}{Integer} \\
        \multicolumn{1}{c||}{\PFibonacci} &
        \multicolumn{1}{l||}{Given a positive integer $n$, calculate the Fibonacci series until the number $n$.} &
        \multicolumn{1}{c}{Integer} &
        \multicolumn{1}{c}{List of integers} \\
        \multicolumn{1}{c||}{\PDivisors} &
        \multicolumn{1}{l||}{Given a positive integer $n$, find its number of divisors that are divisible by $3$.} &
        \multicolumn{1}{c}{Integer} &
        \multicolumn{1}{c}{Integer} \\
        \multicolumn{1}{c||}{\PPalindrome} &
        \multicolumn{1}{l||}{Given a string $S$, check if it is palindrome or not.} &
        \multicolumn{1}{c}{String} &
        \multicolumn{1}{c}{Boolean} \\
        \multicolumn{1}{c||}{\PMergeStrings} &
        \multicolumn{1}{l||}{Given two strings $S_1$ and $S_2$, merge them alternatively.} &
        \multicolumn{1}{c}{Two strings} &
        \multicolumn{1}{c}{String}  \\
        \bottomrule
    \end{tabular}
    }
    \caption{Five introductory Python programming problems used in our work.}
    \label{fig.problemsetup.problems}
\end{figure*}

\begin{figure*}[t!]
\centering
    \scalebox{0.78}{
    \setlength\tabcolsep{2.0pt}
    \renewcommand{\arraystretch}{1.4}
    \begin{tabular}{p{3.5cm}p{3.5cm}||p{9cm}||p{3.5cm}}
        \toprule
        \multicolumn{1}{c}{\textbf{Problem ID}} &
        \multicolumn{1}{c||}{\textbf{Program ID}} &
        \multicolumn{1}{c||}{\textbf{Description of Bug(s)}} &
        \multicolumn{1}{r}{\textbf{\#Lines}} \\
        \midrule
        \multicolumn{1}{c}{\PGCD} &
        \multicolumn{1}{c||}{\BPaGCD} &
        \multicolumn{1}{l||}{`range' function is misused. Time complexity is violated.} &
        \multicolumn{1}{r}{04} \\
        \multicolumn{1}{c}{\PGCD} &        
        \multicolumn{1}{c||}{\BPbGCD} &
        \multicolumn{1}{l||}{`range' function is misused. Time complexity and space complexity are violated.} &
        \multicolumn{1}{r}{17} \\ 
        \multicolumn{1}{c}{\PGCD} &
        \multicolumn{1}{c||}{\BPcGCD} &        
        \multicolumn{1}{l||}{Implementation of the Euclidean algorithm is incorrect.} &
        \multicolumn{1}{r}{07} \\
        \multicolumn{1}{c}{\PGCD} &        
        \multicolumn{1}{c||}{\BPdGCD} &
        \multicolumn{1}{l||}{Wrong arguments are passed to the recursive function call.} &
        \multicolumn{1}{r}{04} \\
        \multicolumn{1}{c}{\PGCD} &        
        \multicolumn{1}{c||}{\BPeGCD} &
        \multicolumn{1}{l||}{A wrong variable is returned.} &
        \multicolumn{1}{r}{06} \\
        \midrule
        \multicolumn{1}{c}{\PFibonacci} &        
        \multicolumn{1}{c||}{\BPaFibonacci} &
        \multicolumn{1}{l||}{Problem description misunderstood.} &
        \multicolumn{1}{r}{09} \\
        \multicolumn{1}{c}{\PFibonacci} &        
        \multicolumn{1}{c||}{\BPbFibonacci} &
        \multicolumn{1}{l||}{Time complexity is violated.} &
        \multicolumn{1}{r}{11} \\      
        \multicolumn{1}{c}{\PFibonacci} &        
        \multicolumn{1}{c||}{\BPcFibonacci} &
        \multicolumn{1}{l||}{Numbers at the end of series are missing.} &
        \multicolumn{1}{r}{07} \\
        \multicolumn{1}{c}{\PFibonacci} &        
        \multicolumn{1}{c||}{\BPdFibonacci} &
        \multicolumn{1}{l||}{Numbers at the beginning of series are missing. Problem description misunderstood.} &
        \multicolumn{1}{r}{15} \\
        \multicolumn{1}{c}{\PFibonacci} &        
        \multicolumn{1}{c||}{\BPeFibonacci} &
        \multicolumn{1}{l||}{Numbers at the beginning of series are missing.} &
        \multicolumn{1}{r}{15} \\
        \midrule
        \multicolumn{1}{c}{\PDivisors} &        
        \multicolumn{1}{c||}{\BPaDivisors} &
        \multicolumn{1}{l||}{Time complexity is violated.} &
        \multicolumn{1}{r}{06} \\
        \multicolumn{1}{c}{\PDivisors} &        
        \multicolumn{1}{c||}{\BPbDivisors} &
        \multicolumn{1}{l||}{Non-divisors are also counted.} &
        \multicolumn{1}{r}{09} \\
        \multicolumn{1}{c}{\PDivisors} &        
        \multicolumn{1}{c||}{\BPcDivisors} &
        \multicolumn{1}{l||}{There is an off-by-one error.} &
        \multicolumn{1}{r}{13} \\
        \multicolumn{1}{c}{\PDivisors} &        
        \multicolumn{1}{c||}{\BPdDivisors} &
        \multicolumn{1}{l||}{Some valid divisors are not counted.} &
        \multicolumn{1}{r}{10} \\
        \multicolumn{1}{c}{\PDivisors} &        
        \multicolumn{1}{c||}{\BPeDivisors} &
        \multicolumn{1}{l||}{Divisors larger than $\sqrt{n}$ are not considered.} &
        \multicolumn{1}{r}{08} \\
        \midrule        
        \multicolumn{1}{c}{\PPalindrome} &        
        \multicolumn{1}{c||}{\BPaPalindrome} &
        \multicolumn{1}{l||}{There is an issue with string indexing. Return type is incorrect.} &
        \multicolumn{1}{r}{09} \\
        \multicolumn{1}{c}{\PPalindrome} &        
        \multicolumn{1}{c||}{\BPbPalindrome} &
        \multicolumn{1}{l||}{All strings with odd lengths are regarded as non-palindrome.} &
        \multicolumn{1}{r}{06} \\
        \multicolumn{1}{c}{\PPalindrome} &        
        \multicolumn{1}{c||}{\BPcPalindrome} &
        \multicolumn{1}{l||}{There is a mistake in the algorithm.} &
        \multicolumn{1}{r}{04} \\
        \multicolumn{1}{c}{\PPalindrome} &        
        \multicolumn{1}{c||}{\BPdPalindrome} &
        \multicolumn{1}{l||}{`return' keyword is missing. Space complexity is violated.} &
        \multicolumn{1}{r}{07} \\
        \multicolumn{1}{c}{\PPalindrome} &        
        \multicolumn{1}{c||}{\BPePalindrome} &
        \multicolumn{1}{l||}{There is a misconception regarding mutability of lists.} &
        \multicolumn{1}{r}{15} \\
        \midrule
        \multicolumn{1}{c}{\PMergeStrings} &        
        \multicolumn{1}{c||}{\BPaMergeStrings} &
        \multicolumn{1}{l||}{Indentation of a statement is incorrect.} &
        \multicolumn{1}{r}{14} \\
        \multicolumn{1}{c}{\PMergeStrings} &        
        \multicolumn{1}{c||}{\BPbMergeStrings} &
        \multicolumn{1}{l||}{There is a mistake regarding lexicographical ordering of strings.} &
        \multicolumn{1}{r}{17} \\
        \multicolumn{1}{c}{\PMergeStrings} &        
        \multicolumn{1}{c||}{\BPcMergeStrings} &
        \multicolumn{1}{l||}{There is a mistake regarding ordering of the merging strings.} &
        \multicolumn{1}{r}{31} \\ 
        \multicolumn{1}{c}{\PMergeStrings} &        
        \multicolumn{1}{c||}{\BPdMergeStrings} &
        \multicolumn{1}{l||}{An if-elif-else statement is misused.} &
        \multicolumn{1}{r}{16} \\
        \multicolumn{1}{c}{\PMergeStrings} &        
        \multicolumn{1}{c||}{\BPeMergeStrings} &
        \multicolumn{1}{l||}{There is a issue regarding the slicing of strings.} &
        \multicolumn{1}{r}{08} \\
        \bottomrule
    \end{tabular}
    }
    \caption{\looseness-1Real-world buggy programs used in our work. These programs are based on publicly accessible submissions from  \emph{geeksforgeeks.org} platform~\cite{geeksforgeeks} and are picked to capture different types of bugs for each problem. The last column, titled ``\#Lines'', indicates the number of lines in the buggy program (not counting the lines that are part of the given template).}   
    \label{fig.problemsetup.programs}
\end{figure*}

\begin{figure}[h!]
    \centering
    \begin{subfigure}[b]{0.87\textwidth}
        \centering
        \scalebox{1.04}{
            \renewcommand{\arraystretch}{1}
            \begin{tabular}{|p{1\linewidth}|}
                \hline
                \multicolumn{1}{|p{0.85\linewidth}|}{
\begin{lstlisting}[basicstyle=\fontsize{7.5}{7}\ttfamily,xleftmargin=1.65em,belowskip=-0.7em,aboveskip=-0.2em,
                    linebackgroundcolor={
                    \ifnum\value{lstnumber}>3
                        \ifnum\value{lstnumber}<15
                            \color{CodeHighlight}
                        \fi
                    \fi
                    }]
#User function Template for python3
class Solution:
    def nFibonacci(self,N):
        # base cases
        if N == 0: return [0]
        if N == 1: return [0, 1, 1]
        # create the fibonacci sequence
        fib_sequence = [0, 1]
        while True:
            next_value = fib_sequence[-1] + fib_sequence[-2]
            if next_value > N: 
                break
            fib_sequence.append(next_value)
        return fib_sequence

# Driver Code Starts
#Initial Template for Python 3
if __name__=='__main__':
    t=int(input())
    for _ in range(t):
        N=int(input())
        ob=Solution()
        ans=ob.nFibonacci(N)
        for i in ans:
            print(i,end=" ")
        print()
# } Driver Code Ends
\end{lstlisting}}\\
                \hline
            \end{tabular}
        }
    \end{subfigure}
    \caption{A solution program generated by  \TechGPTFour{} for \PFibonacci{}. The highlighted lines are generated by \TechGPTFour{}; the rest, including the \emph{Driver Code}, is part of the solution template on \emph{geeksforgeeks.org} platform~\cite{geeksforgeeks}. This template, along with the problem description, is given in the prompt.}
    \label{fig.problemsetup.solutions.Fibonacci}
\end{figure}

\paragraph{Methods evaluated.} We evaluate three methods in our work: (a) \TechChatGPT{} that uses OpenAI's ChatGPT (based on GPT-3.5) as its LLM via web platform~\cite{ChatGPT,ChatGPTWeb}; (b) \TechGPTFour{} that uses OpenAI's GPT-4 as its LLM via web platform~\cite{GPT4,GPT4Web}; (c) \TechTutor{} that corresponds to human experts with experience in Python programming and tutoring introductory programming classes. Prompts used to interact with LLMs are provided in the subsequent sections for different scenarios. The information in these prompts also serves as instructions for human experts that are part of the \TechTutor{} method; these experts can naturally draw on their own experiences and use additional resources---e.g., debugger, web, or course materials---similar to how a typical human tutor/educator would work on these scenarios in real-world settings. Next, we describe the interaction process with these models and outputs for evaluation. For a given method and scenario,  we have $25$ total instances for evaluation, comprising a problem and program ($5\times5$ instances). For each instance, \TechChatGPT{} and \TechGPTFour{} perform $n_{\text{\TechChatGPT{}}}=n_{\text{\TechGPTFour{}}}=1$ query to their corresponding LLMs through web platforms to generate one output per instance; \TechTutor{} has $n_{\text{\TechTutor{}}}=2$ human experts that independently generate two outputs per instance.\footnote{We note that GPT-4 currently has a cap of $25$ messages every $3$ hours~\cite{GPT4Web}. In future studies, it would be useful to scale up the evaluation process by increasing $n_{\text{\TechChatGPT{}}}$, $n_{\text{\TechGPTFour{}}}$, and $n_{\text{\TechTutor{}}}$.} We describe further scenario-specific details in the subsequent sections.

\paragraph{Metrics and evaluation process.} We will introduce scenario-specific performance metrics in the subsequent sections. We have $n_{\text{\Evaluators{}}}=2$ human evaluators who provide annotations to assess the quality of generated output for each instance w.r.t. corresponding performance metrics. In our evaluation, this set of $n_{\text{\Evaluators{}}}=2$ human evaluators is same as $n_{\text{\TechTutor{}}}=2$ human experts that are part of the \TechTutor{} method. More concretely, each of the $n_{\text{\Evaluators{}}}$ human evaluators independently annotates the quality of generated outputs for \TechChatGPT{}, \TechGPTFour{}, and \TechTutor{} (only for the $n_{\text{\TechTutor{}}}\!-\!1$ human experts by excluding the evaluator themselves). Then, for each method, results are first aggregated across $25$ instances or across $5$ instances when reporting problem-specific results. Finally, we aggregate results across $n_{\text{\Evaluators{}}}$ human evaluators and report averaged results as \emph{mean} (\emph{stderr}). We provide scenario-specific details in the subsequent sections.

\paragraph{Remark.} Since we want LLMs to play the role of experienced digital tutors and assistants, a natural question is whether they can solve five problems used in the evaluation, i.e., generate correct solution programs. Before evaluating \TechChatGPT{} and \TechGPTFour{} on different programming education scenarios, we checked their problem-solving capabilities by querying them with suitable prompts consisting of a problem description along with a solution template as input and instruction to generate a solution program as output. \TechGPTFour{} was able to solve all five problems; Figure~\ref{fig.problemsetup.solutions.Fibonacci} above and Figures~\ref{fig.problemsetup.solutions.MergeStrings}--\ref{fig.problemsetup.solutions.Palindrome} in Appendix~\ref{app.sec.problemsetup} show solution programs generated by \TechGPTFour{} for these problems. \TechChatGPT{} was able to solve four out of five problems; it consistently failed on \PFibonacci{} across multiple queries.

%

\section{Program Repair Scenario}\label{sec.programrepair}
This section is dedicated to the programming education scenario of \emph{program repair}~\cite{yi2017feasibility,zhang2022repairing,edm23-pyfixv}. This scenario is motivated by an AI-based educational agent acting as a \emph{digital tutor for a student} and providing help by fixing the student's buggy program. Next, we provide details of this scenario's prompt, input-output formats, performance metrics, and results.

\paragraph{Prompt and output generation.} We begin by describing the content provided as input to a method and the desired output content we seek to generate. The input consists of a detailed \emph{problem description} and a student's \emph{buggy program}; the desired output consists of a \emph{fixed program}. Figure~\ref{fig.program_repair.prompt} shows the prompt---with placeholders for the inputs---used to interact with LLMs for \TechChatGPT{} and \TechGPTFour{} methods. The prompt starts with an overview text about the scenario, followed by a detailed problem description and a student's buggy program, and then summarizes the desired output. When interacting with LLMs, we first generate content using this prompt and then manually extract the generated program as the final output for evaluation.

\paragraph{Output quality and performance metrics.} We assess the generated output along several quality attributes and use aggregated results over these quality attributes as performance metrics in our evaluation. \emph{Correct} (binary, $1$ value being better) captures whether the generated program is correct w.r.t. the problem specification; we use automated test suites to check the correctness of a program as mentioned in Footnote~\ref{footnote.problemdescription}. \emph{EditTokens} (non-negative number, lower value being better) captures the token-based edit distance between the generated program and the buggy program.\footnote{Edit-distance between two programs is measured by first tokenizing programs using Pygments library~\cite{pygments} and then computing Levenshtein edit-distance over token strings.} Human evaluators annotate the quality of generated output for each of the $25$ instances; in this particular scenario, human evaluators computed these attributes using automated scripts without requiring manual annotation. 

%

\begin{figure}[t!]
    \centering
    \scalebox{0.96}{
        \setlength\tabcolsep{5pt}
        \renewcommand{\arraystretch}{1.2}
\begin{tabular}{||p{0.99\linewidth}||}
    \hline
    \multicolumn{1}{||c||}{\promptheader{Prompt: Program Repair}} \\ 
    I'm working on a Python programming problem. The current program below is not working well. Can you help in fixing this program with as few changes as possible? Below I first provide the problem description and then the current buggy program.
    \newline
    \newline
    \promptinput{\{problem\_description\}}
    \newline
    \newline
    Buggy Program:
    \newline
    \newline
    \`{}\`{}\`{}\vspace{-2mm}\\
    \promptinput{\{buggy\_program\}}\\
    \`{}\`{}\`{}
    \newline
    \newline
    Can you fix the above buggy program? The code marked as \#Driver Code is correct and should not be modified. Make sure that you make minimal possible changes needed to fix the program.
    \\
    \hline
\end{tabular}
%
    }
    \caption{Prompt for the program repair scenario. This prompt has two placeholders for the problem description and the buggy program.
    }
    \label{fig.program_repair.prompt}
\end{figure}

\begin{figure*}[t!]
\centering
    \begin{subfigure}[b]{1\textwidth}
        \centering
        \scalebox{0.78}{
        \setlength\tabcolsep{10.0pt}
        \renewcommand{\arraystretch}{1.3}        
\begin{tabular}{l||cc}
    \toprule
    \multicolumn{1}{c||}{\textbf{Method}} & \multicolumn{2}{c}{\textbf{(Fixed Program, Buggy Program)}}\\
     & \multicolumn{1}{c}{Correct} & \multicolumn{1}{c}{EditTokens}\\  
    \midrule
    \TechChatGPT & $\ \ 68.0~(0.0)$ & $43.0~(0.0)$ \\
    \TechGPTFour & $\ \ 88.0~(0.0)$ & $36.6~(0.0)$ \\
    {\cellcolor{\tutorcellcolor}\TechTutor} & {\cellcolor{\tutorcellcolor}$100.0~(0.0)$} & {\cellcolor{\tutorcellcolor}$19.0~(1.2)$} \\
    \bottomrule
\end{tabular}


        }
        \vspace{-1mm}
        \caption{}
        \label{fig.program_repair.results.table}
        \vspace{1mm}
    \end{subfigure}
    \\ 
    %
    \begin{subfigure}[b]{1\textwidth}
        \centering
        \includegraphics[height=3.8cm]{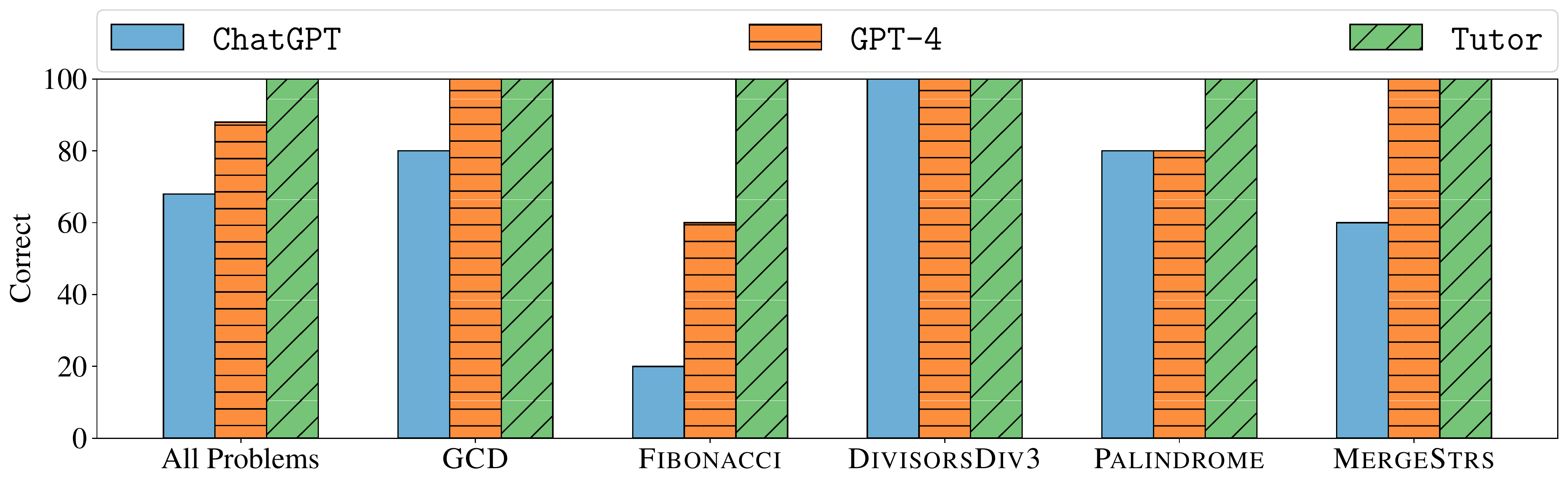}
        \vspace{-2mm}
        \caption{}
        \label{fig.program_repair.results.histogram}
    \end{subfigure}
    \caption{Results for the program repair scenario. \textbf{(a)} Results for various metrics aggregated across all problems. \textbf{(b)} Results for the metric \emph{Correct} separately on five problems. For the metric \emph{Correct}, these aggregated results are reported in terms of \%. Details are in Section~\ref{sec.programrepair}.}
    \label{fig.program_repair.results}
    \vspace{-2.5mm}    
\end{figure*}

\paragraph{Results.} Figure~\ref{fig.program_repair.results.table} provide results for various metrics aggregated across all problems, and Figure~\ref{fig.program_repair.results.histogram} for the metric \emph{Correct} separately on five problems. These aggregated results for the metric \emph{Correct} are reported in terms of \%. Next, we summarize some of the key findings. First, results in Figure~\ref{fig.program_repair.results.table} for the metric \emph{Correct} highlight that \TechGPTFour{} ($88.0$) substantially improves up on \TechChatGPT{} ($68.0$) and comes close to the performance of \TechTutor{} ($100.0$). However, in terms of the metric \emph{EditTokens}, \TechGPTFour{} ($36.6$) does a lot more edits when fixing buggy programs in contrast to that made by \TechTutor{} ($19.0$). Second, results in Figure~\ref{fig.program_repair.results.histogram} highlight that these findings are generally consistent across all five problems for the metric \emph{Correct}; the gap in the performance of \TechGPTFour{} vs. \TechTutor{} is worst on \PFibonacci{} for this scenario. In Appendix~\ref{app.sec.programrepair}, we provide an illustrative example showing the outputs of different methods.

%

\section{Hint Generation Scenario}\label{sec.hintgeneration}
This section is dedicated to the programming education scenario of \emph{hint generation}~\cite{leinonen23sigcse,edm23-pyfixv}. This scenario is motivated by an AI-based educational agent acting as a \emph{digital tutor for a student} and providing help via hints to resolve current issues in the student's buggy program. Next, we provide details of this scenario's prompt, input-output formats, performance metrics, and results.

\paragraph{Prompt and output generation.} We begin by describing the content provided as input to a method and the desired output content we seek to generate. The input consists of a detailed \emph{problem description} and a student's \emph{buggy program}; the desired output consists of a \emph{hint} and an \emph{explanation} that provides the reasoning behind the generated hint. Figure~\ref{fig.hint_generation.prompt} shows the prompt---with placeholders for the inputs---used to interact with LLMs for \TechChatGPT{} and \TechGPTFour{} methods.
When interacting with LLMs, we first generate content using this prompt and then manually extract the generated hint and explanation as the final output used for evaluation.

\paragraph{Output quality and performance metrics.} We assess the generated output along several quality attributes and use aggregated results over these quality attributes as performance metrics in our evaluation. All attributes for this scenario are binary, with a value of $1$ being better. \emph{HCorrect} (binary) captures whether the generated hint provides correct information for resolving issues in the student's buggy program. \emph{HInformative} (binary) captures whether the generated hint provides useful information to help the student resolve bug(s); this attribute is set to $0$ by default when the hint is incorrect. \emph{HConceal} (binary) captures that the information in the generated hint is not too detailed, so the student would also have to reason about implementing the fixes; this attribute is set to $0$ by default when the hint is incorrect. \emph{HComprehensible} (binary) captures whether the generated hint is easy to understand, presented in a readable format, and doesn't contain redundant information. \emph{HOverall} (binary) is $1$, i.e., good quality, only if the generated hint satisfies \emph{all} the four quality attributes mentioned above. \emph{ECorrect} (binary) captures whether the generated explanation contains correct reasoning behind the generated hint; this attribute is set to $0$ by default when the hint is incorrect. \emph{Overall} (binary) is $1$ only when \emph{both} the \emph{HOverall} and \emph{ECorrect} attributes are $1$. Human evaluators annotate the quality of generated output for each of the $25$ instances; in this scenario, these attributes require manual annotation (in contrast to automated annotation for the program repair scenario).
%

%

\begin{figure}[t!]
    \centering
    \scalebox{0.96}{
        \setlength\tabcolsep{5pt}
        \renewcommand{\arraystretch}{1.2}
\begin{tabular}{||p{0.99\linewidth}||}
    \hline
    \multicolumn{1}{||c||}{\promptheader{Prompt: Hint Generation}} \\ 
    I'm working on a Python programming problem. The current program below is not working well. Can you help by giving a hint? Below I first provide the problem description and then the current buggy program.
    \newline
    \newline
    \promptinput{\{problem\_description\}}
    \newline
    \newline
    Buggy Program:
    \newline
    \newline
    \`{}\`{}\`{}\vspace{-2mm}\\
    \promptinput{\{buggy\_program\}}\\
    \`{}\`{}\`{}
    \newline
    \newline
    (1) Can you describe the bug(s) in this program and the required fixes?
    \newline
    \newline
    (2) Can you provide a concise single-sentence hint about one bug in this program? The hint should not be too detailed as I want to think about the fixes by myself. However, the hint should not be too abstract, as I need some help.
    \\
    \hline
\end{tabular}

%
    }
    \caption{Prompt for the hint generation scenario. This prompt has two placeholders for the problem description and the buggy program.
    }
    \label{fig.hint_generation.prompt}
\end{figure}

\begin{figure*}[t!]
\centering
    \begin{subfigure}[b]{1\textwidth}
        \centering
        \scalebox{0.78}{
        \setlength\tabcolsep{4pt}
        \renewcommand{\arraystretch}{1.3}
\begin{tabular}{l||c|ccccc|c}
    \toprule
    \multicolumn{1}{c||}{\textbf{Method}} & \multicolumn{1}{c|}{\textbf{(Hint, Explanation)}} & \multicolumn{5}{c|}{\textbf{Hint}} & \multicolumn{1}{c}{\textbf{Explanation}}\\
     & \multicolumn{1}{c|}{Overall} & \multicolumn{1}{c}{HOverall} & \multicolumn{1}{c}{HCorrect} & \multicolumn{1}{c}{HInformative} & \multicolumn{1}{c}{HConceal} & \multicolumn{1}{c|}{HComprehensible} & \multicolumn{1}{c}{ECorrect}\\  
    \midrule
    \TechChatGPT & $18.0~(\ \ 6.0)$ & $22.0~(\ \ 6.0)$ & $50.0~(10.0)$ & $38.0~(\ \ 6.0)$ & $38.0~(10.0)$ & $94.0~(\ \ 2.0)$ & $40.0~(\ \ 8.0)$ \\
    \TechGPTFour & $66.0~(10.0)$ & $70.0~(10.0)$ & $74.0~(10.0)$ & $74.0~(10.0)$ & $72.0~(\ \ 8.0)$ & $96.0~(\ \ 4.0)$ & $70.0~(10.0)$ \\
    {\cellcolor{\tutorcellcolor}\TechTutor} & {\cellcolor{\tutorcellcolor}$92.0~(\ \ 4.0)$} & {\cellcolor{\tutorcellcolor}$92.0~(\ \ 4.0)$} & {\cellcolor{\tutorcellcolor}$94.0~(\ \ 6.0)$} & {\cellcolor{\tutorcellcolor}$94.0~(\ \ 6.0)$} & {\cellcolor{\tutorcellcolor}$94.0~(\ \ 6.0)$} & {\cellcolor{\tutorcellcolor}$98.0~(\ \ 2.0)$} & {\cellcolor{\tutorcellcolor}$94.0~(\ \ 6.0)$} \\
    \bottomrule
\end{tabular}

        }
        \vspace{-1mm}        
        \caption{}
        \label{fig.hint_generation.results.table}
        \vspace{1mm}
    \end{subfigure}
    \\
    %
    \begin{subfigure}[b]{1\textwidth}
        \centering
        \includegraphics[height=3.8cm]{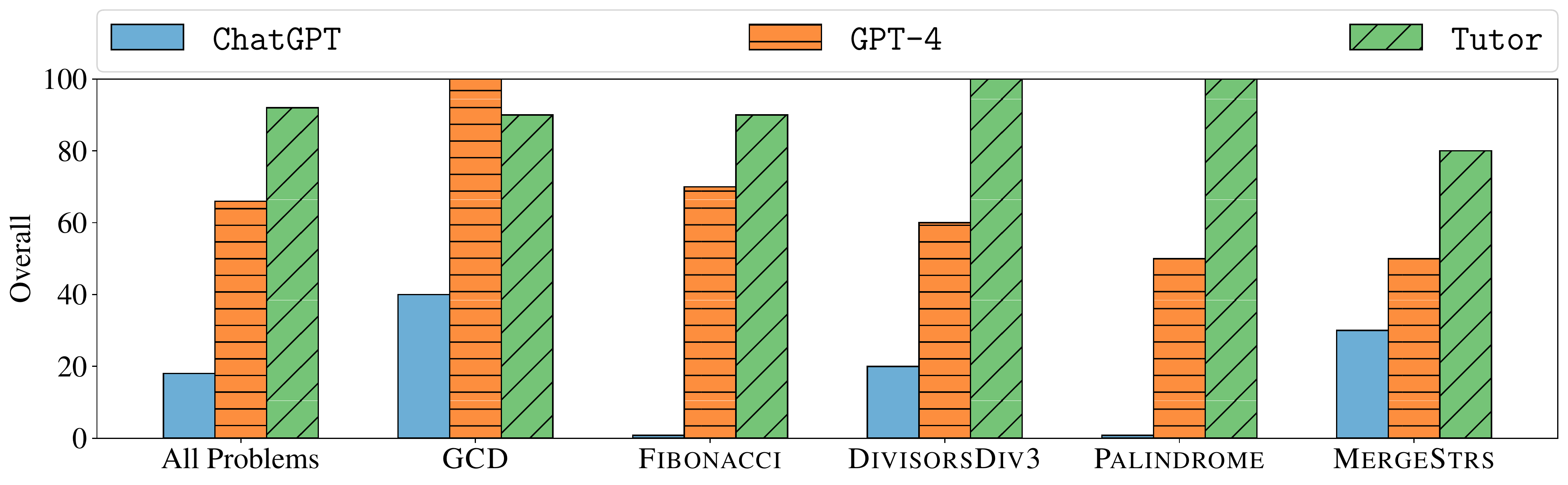}
        \vspace{-2mm}
        \caption{}
        \label{fig.hint_generation.results.histogram}
    \end{subfigure}
    \caption{Results for the hint generation scenario. \textbf{(a)} Results for various metrics aggregated across all problems. \textbf{(b)} Results for the metric \emph{Overall} separately on five problems. For all metrics, these aggregated results are reported in terms of \%. Details are in Section~\ref{sec.hintgeneration}.}
    \label{fig.hint_generation.results}
    \vspace{-2.5mm}    
\end{figure*}

\paragraph{Results.} Figure~\ref{fig.hint_generation.results.table} provide results for various metrics aggregated across all problems, and Figure~\ref{fig.hint_generation.results.histogram} for the metric \emph{Overall} separately on five problems. These aggregated results for various metrics are reported in terms of \%. Next, we summarize some of the key findings. First, results in Figure~\ref{fig.hint_generation.results.table} for the metric \emph{Overall} highlight that \TechGPTFour{} ($66.0$) substantially improves up on \TechChatGPT{} ($18.0$), though there is still a large gap in comparison to the performance of \TechTutor{} ($92.0$). Combining this with results on metrics \emph{HOverall} and \emph{ECorrect}, we can see that \TechGPTFour{}'s detailed reasoning is usually correct whenever it manages to generate a good quality hint. Second, results in Figure~\ref{fig.hint_generation.results.histogram} highlight that these findings are generally consistent across all five problems for the metric \emph{Overall}; the gap in the performance of \TechGPTFour{} vs. \TechTutor{} is worst on \PPalindrome{} for this scenario. Interestingly, \TechGPTFour{}'s performance on \PGCD{} is slightly better than that of \TechTutor{}. In Appendix~\ref{app.sec.hintgeneration}, we provide an illustrative example to qualitatively show the outputs of different methods.

%

\section{Grading Feedback Scenario}\label{sec.gradingfeedback}
This section is dedicated to the programming education scenario of \emph{grading feedback}~\cite{DBLP:journals/corr/abs-2106-07340,DBLP:conf/aied/FunayamaSMMSI22}. This scenario is motivated by an AI-based educational agent acting as a \emph{digital assistant for an educator} and providing assistance by grading students' programs w.r.t. a given rubric. Next, we provide details of this scenario's prompt, input-output formats, performance metrics, and results.

\begin{figure}[t!]
    \centering
    \scalebox{0.96}{
        \setlength\tabcolsep{5pt}
        \renewcommand{\arraystretch}{1.2}        
\begin{tabular}{||p{0.99\linewidth}||}
    \hline
    \multicolumn{1}{||c||}{\promptheader{Prompt: Grading Feedback}} \\ 
    I have to grade a student's program for a Python programming problem. Can you help in grading this student’s program according to a given rubric? Below I first provide the problem description, the student’s program, and then the grading rubric.
    \newline
    \newline
    \promptinput{\{problem\_description\}}
    \newline
    \newline
    Student's Program:
    \newline
    \newline
    \`{}\`{}\`{}\vspace{-2mm}\\
    \promptinput{\{student\_program\}}\\
    \`{}\`{}\`{}
    \newline
    \newline
    Grading Rubric:
    \newline
    \newline
    The grading rubric is divided into five dimensions to assess different aspects of the program. The maximum possible points for a program is $100$.
    \newline
    \newline
    1. Program format ($10$ points)
    \newline
    - The program implements the \promptinput{\{function\_name\}} function with the correct input parameters and return type as specified in the problem description. The possible points along this rubric dimension are $10$ or $0$.
    \newline
    \newline
    2. Time complexity ($15$ points)
    \newline
    - The program meets the expected time complexity of \promptinput{\{time\_complexity\}}. The possible points along this rubric dimension are $15$ or $0$.
    \newline
    \newline
    3. Space complexity ($15$ points)
    \newline
    - The program meets the expected auxiliary space complexity of \promptinput{\{space\_complexity\}}. The possible points along this rubric dimension are $15$ or $0$.
    \newline
    \newline
    4. Correctness for general inputs ($30$ points)
    \newline
    - The program outputs the correct results for most of the inputs. We give full points as long as the program works correctly for general inputs, i.e., excluding specific inputs that are edge cases or cause issues with time/space complexity. The possible points along this rubric dimension are $30$ or $0$.
    \newline
    \newline
    5. Correctness for edge cases ($30$ points)
    \newline
    - The program outputs the correct results for specific inputs that are edge cases. We subtract 10 points for each failing type of edge case. We give 0 points when there are more than three types of failing edge cases. The possible points along this rubric dimension are $30$, $20$, $10$ or $0$.
    \newline
    \newline
    Can you grade the student's program according to the above five-dimensional rubric and also provide total points scored by the student's program? The code marked as \#Driver Code is correct and should not be considered for grading.
    \\
    \hline
\end{tabular}
    }
    \caption{\looseness-1Prompt for the grading feedback scenario. This prompt has two placeholders for the problem description and the student's program, and three placeholders for problem-specific details in the rubric.
    }
    \label{fig.grading_feedback.prompt}
    \vspace{-4.5mm}
\end{figure}

\begin{figure*}[t!]
\centering
    \begin{subfigure}[b]{1\textwidth}
        \centering
        \scalebox{0.78}{
        \setlength\tabcolsep{1.8pt}
        \renewcommand{\arraystretch}{1.3}
\begin{tabular}{l||c|cccccc}
    \toprule
    \multicolumn{1}{c||}{\textbf{Method}} & \multicolumn{7}{c}{\textbf{Match and Difference (D-) in Points Across Grading Rubric}}\\
     & \multicolumn{1}{c|}{Match} & \multicolumn{1}{c}{D-Total} & \multicolumn{1}{c}{D-ProgramFormat} & \multicolumn{1}{c}{D-TimeComplexity} & \multicolumn{1}{c}{D-SpaceComplexity} & \multicolumn{1}{c}{D-CorrectGeneral} & \multicolumn{1}{c}{D-CorrectEdge} \\  
    \midrule
    \TechChatGPT & $10.0~(6.0)$ & $36.7~(1.3)$ & $\ \ 0.8~(0.4)$ & $\ \ 0.6~(0.6)$ & $\ \ 2.7~(0.9)$ & $18.0~(1.2)$ & $18.2~(0.6)$ \\
    \TechGPTFour & $16.0~(4.0)$ & $23.5~(2.9)$ & $\ \ 1.2~(0.4)$ & $\ \ 1.8~(0.6)$ & $\ \ 3.3~(0.9)$ & $\ \ 8.4~(1.2)$ & $11.0~(0.6)$ \\
    {\cellcolor{\tutorcellcolor}\TechTutor} & {\cellcolor{\tutorcellcolor}$52.0~(0.0)$} & {\cellcolor{\tutorcellcolor}$\ \ 9.0~(0.0)$} & {\cellcolor{\tutorcellcolor}$\ \ 0.8~(0.0)$} & {\cellcolor{\tutorcellcolor}$\ \ 1.2~(0.0)$} & {\cellcolor{\tutorcellcolor}$\ \ 1.8~(0.0)$} & {\cellcolor{\tutorcellcolor}$\ \ 2.4~(0.0)$} & {\cellcolor{\tutorcellcolor}$\ \ 3.6~(0.0)$} \\
    \bottomrule
\end{tabular}

        }
        \vspace{-1mm}        
        \caption{}
        \label{fig.grading_feedback.results.table}
        \vspace{1mm}
    \end{subfigure}
    \\
    %
    \begin{subfigure}[b]{1\textwidth}
        \centering
        \includegraphics[height=3.8cm]{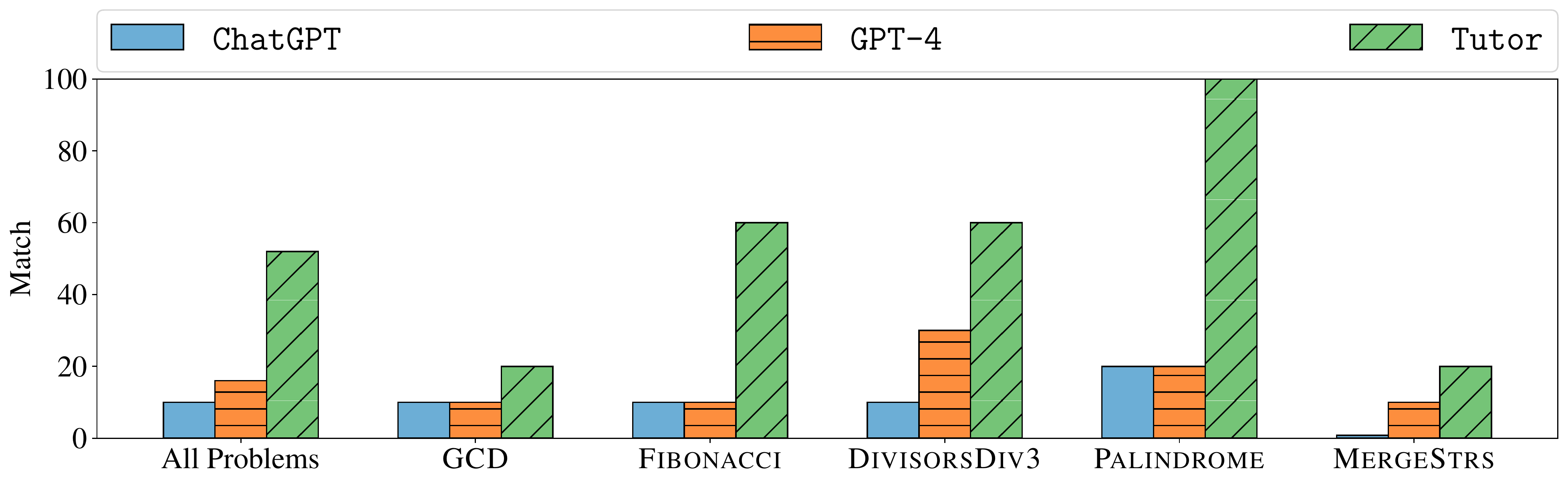}
        \vspace{-2mm}
        \caption{}
        \label{fig.grading_feedback.results.histogram}
    \end{subfigure}
    \caption{Results for the grading feedback scenario. \textbf{(a)} Results for various metrics aggregated across all problems. \textbf{(b)} Results for the metric \emph{Match} separately on five problems. For the metric \emph{Match}, these aggregated results are reported in terms of \%, and higher values correspond to better performance. For the metrics related to difference (D-), lower differences in points correspond to better performance. Details are in Section~\ref{sec.gradingfeedback}.}
    \label{fig.grading_feedback.results}
    \vspace{-3.5mm}    
\end{figure*}

\paragraph{Prompt and output generation.} We begin by describing the content provided as input to a method and the desired output content we seek to generate. The input consists of a detailed \emph{problem description}, a \emph{student's program} with bugs, and a \emph{grading rubric}; the desired output consists of \emph{grading points} w.r.t. the grading rubric. Figure~\ref{fig.grading_feedback.prompt} shows the prompt---with placeholders for the inputs---used to interact with LLMs for \TechChatGPT{} and \TechGPTFour{} methods. 
When interacting with LLMs, we first generate content using this prompt and then manually extract the grading points w.r.t. the rubric as the final output for evaluation.\footnote{LLMs typically also generate explanations along with grading points; we do not evaluate the quality of these explanations. In Appendix~\ref{app.sec.gradingfeedback}, we will provide an illustrative example to show some of these explanations.}

\paragraph{Output quality and performance metrics.} We assess the generated output along several quality attributes and use aggregated results over these quality attributes as performance metrics in our evaluation. \emph{Match} (binary, $1$ value being better) captures the exact match between the points w.r.t. grading rubric provided by the method and that provided by the human evaluator during annotation.
\emph{D-Total} (non-negative number, lower value being better) captures the absolute difference between the total points provided by the method and that provided by the human evaluator during annotation. Moreover, we consider attributes corresponding to the grading rubric specified in the prompt; here, we have used a five-dimensional rubric with rubric dimensions of program format, time complexity, space complexity, correctness for general inputs, and correctness for edge cases. For this rubric, we have five additional attributes, namely, \emph{D-ProgramFormat}, \emph{D-TimeComplexity}, \emph{D-SpaceComplexity}, \emph{D-CorrectGeneral}, and \emph{D-CorrectEdge}, that are variants of \emph{D-Total} for computing absolute differences along specific rubric dimensions.

\paragraph{Results.} Figure~\ref{fig.grading_feedback.results.table} provide results for various metrics aggregated across all problems, and Figure~\ref{fig.grading_feedback.results.histogram} for the metric \emph{Match} separately on five problems. These aggregated results for the metric \emph{Match} are reported in terms of \%, and higher values correspond to better performance. Next, we summarize some of the key findings. First, results in Figure~\ref{fig.grading_feedback.results.table} for the metric \emph{Match} highlight that even though \TechGPTFour{} ($16.0$) improves up on \TechChatGPT{} ($10.0$), it still performs substantially worse in comparison to the performance of \TechTutor{} ($52.0$). Further, results on various metrics suggest that the gap in the performance of \TechGPTFour{} and \TechTutor{} is worse on the metric \emph{D-CorrectEdge} that requires more in-depth reasoning about bugs in terms of failing edge cases. Second, results in Figure~\ref{fig.grading_feedback.results.histogram} highlight that these findings are generally consistent across all five problems for the metric \emph{Match}; the gap in the performance of \TechGPTFour{} vs. \TechTutor{} is worst on \PFibonacci{} and \PPalindrome{}. In Appendix~\ref{app.sec.gradingfeedback}, we provide an illustrative example to qualitatively show the outputs of different methods.

%
%

\vspace{-0.5mm}
\section{Pair Programming Scenario}\label{sec.pairprogramming}
\vspace{-0.5mm}
\looseness-1This section is dedicated to the programming education scenario of \emph{pair programming}~\cite{CopilotWeb,DBLP:journals/corr/abs-2210-14306,DBLP:conf/icse/Imai22,DBLP:journals/corr/abs-2306-05153}. This scenario is motivated by an AI-based educational agent acting as a \emph{digital peer for a student} and collaborating via completing an incomplete/partial program written by the student. In contrast to the program repair scenario where the input is a complete student's program with bugs, here the input is an incomplete student's program (e.g., half-done program) that the AI agent is expected to complete. Next, we provide details of this scenario's prompt, input-output formats, performance metrics, and results.
%

\paragraph{Prompt and output generation.} We begin by describing the content provided as input to a method and the desired output content we seek to generate. The input consists of a detailed \emph{problem description} and a student's \emph{partial program}; the desired output consists of a \emph{completed program}.\footnote{In our evaluation, we obtain these partial programs from $25$ buggy programs by removing the second half of the program in terms of the number of lines (only considering lines that are not part of the given template). Importantly, note that the partial program could have bugs or could be using some wrong algorithmic procedure.\label{footnote.pairprogramming}} Figure~\ref{fig.pair_programming.prompt} shows the prompt used to interact with LLMs for \TechChatGPT{} and \TechGPTFour{} methods. 
When interacting with LLMs, we first generate content using this prompt and then manually extract the completed program as the final output for evaluation.

\begin{figure}[t!]
    \centering
    \scalebox{0.96}{
        \setlength\tabcolsep{5pt}
        \renewcommand{\arraystretch}{1.2}
\begin{tabular}{||p{0.99\linewidth}||}
    \hline
    \multicolumn{1}{||c||}{\promptheader{Prompt: Pair Programming}} \\ 
    I'm working on a Python programming problem. I have written a part of the program. Can you help in completing this partial program by adding new lines of code? You should make as few changes as possible to already written lines in the partial program. Below I first provide the problem description and then my partial program.
    \newline
    \newline
    \promptinput{\{problem\_description\}}
    \newline
    \newline
    Partial Program:
    \newline
    \newline
    \`{}\`{}\`{}\vspace{-2mm}\\
    \promptinput{\{partial\_program\}}\\
    \`{}\`{}\`{}
    \newline
    \newline
    Can you complete the above partial program? The code marked as \#Driver Code is correct and should not be modified. Make sure that you make minimal possible changes to already written lines in my partial program.
    \\
    \hline
\end{tabular}

    }
    \caption{Prompt for the pair programming scenario. This prompt has two placeholders for the problem description and the partial program.
    }
    \label{fig.pair_programming.prompt}
\end{figure}

\begin{figure*}[t!]
\centering
    \begin{subfigure}[b]{1\textwidth}
        \centering
        \scalebox{0.78}{
        \setlength\tabcolsep{10.0pt}
        \renewcommand{\arraystretch}{1.3}
\begin{tabular}{l||cccc}
    \toprule
    \multicolumn{1}{c||}{\textbf{Method}} & \multicolumn{4}{c}{\textbf{(Completed Program, Partial Program)}}\\
     & \multicolumn{1}{c}{Overall} & \multicolumn{1}{c}{Correct} & \multicolumn{1}{c}{ContextKept} & \multicolumn{1}{c}{EditLines} \\  
    \midrule
    %
    %
    %
    \TechChatGPT & $38.0~(\ \ 2.0)$ & $\ \ 52.0~(0.0)$ & $86.0~(\ \ 2.0)$ & $6.3~(0.0)$ \\
    \TechGPTFour & $64.0~(\ \ 0.0)$ & $\ \ 92.0~(0.0)$ & $72.0~(\ \ 0.0)$ & $7.7~(0.0)$ \\
    {\cellcolor{\tutorcellcolor}\TechTutor} & {\cellcolor{\tutorcellcolor}$82.0~(10.0)$} & {\cellcolor{\tutorcellcolor}$100.0~(0.0)$} & {\cellcolor{\tutorcellcolor}$82.0~(10.0)$} & {\cellcolor{\tutorcellcolor}$6.0~(0.1)$} \\
    \bottomrule
\end{tabular}
        }
        \vspace{-1mm}
        \caption{}
        \label{fig.pair_programming.results.table}
        \vspace{1mm}
    \end{subfigure}
    \\
    %
    \begin{subfigure}[b]{1\textwidth}
        \centering
        \includegraphics[height=3.8cm]{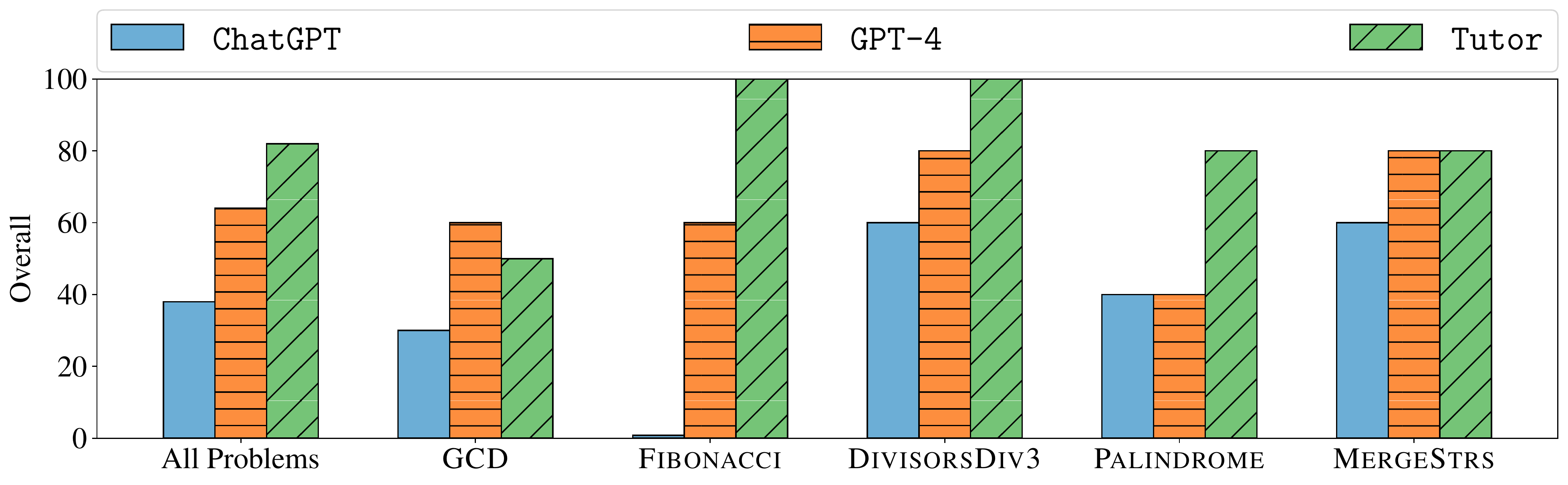}
        \vspace{-2mm}
        \caption{}
        \label{fig.pair_programming.results.histogram}
    \end{subfigure}
    \caption{\looseness-1Results for the pair programming scenario. \textbf{(a)} Results for various metrics aggregated across all problems. \textbf{(b)} Results for the metric \emph{Overall} separately on five problems. For metrics \emph{Correct}, \emph{ContextKept}, and \emph{Overall}, aggregated results are reported in terms of \%. Details are in Section~\ref{sec.pairprogramming}.}
    \label{fig.pair_programming.results}
    \vspace{-2.5mm}    
\end{figure*}

\paragraph{Output quality and performance metrics.} We assess the generated output along several quality attributes and use aggregated results over these quality attributes as performance metrics in our evaluation. \emph{Correct} (binary, $1$ value being better) captures whether the completed program is correct w.r.t. the problem specification; we use automated test suites to check the correctness of a program as mentioned in Footnote~\ref{footnote.problemdescription}. \emph{ContextKept} (binary, $1$ value being better) captures whether the completed program keeps the context of the partial program, e.g., variable names. \emph{EditLines} (non-negative number, lower value being better) captures the line-based edit distance between the completed program and the partial program.\footnote{Edit-distance between two programs is measured based on computing line-diff between them and counting the number of lines that differ.} \emph{Overall} (binary, $1$ value being better) is $1$ only when \emph{both} the \emph{Correct} and \emph{ContextKept} attributes are  $1$. Human evaluators annotate the quality of generated output for each of the $25$ instances. In this scenario, human evaluators computed attributes \emph{Correct} and \emph{EditLines} using automated scripts without requiring manual annotation, and computed the attribute \emph{ContextKept} using manual annotation. 

\paragraph{Results.} Figure~\ref{fig.pair_programming.results.table} provide results for various metrics aggregated across all problems, and Figure~\ref{fig.pair_programming.results.histogram} for the metric \emph{Overall} separately on five problems. These aggregated results for metrics \emph{Correct}, \emph{ContextKept}, and \emph{Overall} are reported in terms of \%. Next, we summarize some of the key findings. First, results in Figure~\ref{fig.pair_programming.results.table} for the metric \emph{Overall} highlight that \TechGPTFour{} ($64.0$) substantially improves up on \TechChatGPT{} ($38.0$) and closed about half the gap in comparison to the performance of \TechTutor{} ($82.0$). However, the results on metrics \emph{ContextKept} and \emph{EditLines} indicate that \TechGPTFour{} has the tendency to make more edits and thereby not keep the context in the partial program provided as input. Second, results in Figure~\ref{fig.pair_programming.results.histogram} highlight that these findings are generally consistent across all five problems for the metric \emph{Overall}; the gap in the performance of \TechGPTFour{} vs. \TechTutor{} is worst on \PFibonacci{} and \PPalindrome{} for this scenario. Interestingly, \TechGPTFour{}'s performance on \PGCD{} is slightly better than that of \TechTutor{}. These results on specific problems of \PFibonacci{}, \PPalindrome{}, and \PGCD{} are aligned with what we observed for the scenarios of program repair (Figure~\ref{fig.program_repair.results.histogram}) and hint generation (Figure~\ref{fig.hint_generation.results.histogram}). In Appendix~\ref{app.sec.pairprogramming}, we provide an illustrative example to qualitatively show the outputs of different methods.

\section{Contextualized Explanation Scenario}\label{sec.contextualizedexplanation}
\looseness-1This section is dedicated to the programming education scenario of \emph{contextualized explanation}~\cite{DBLP:conf/icer/SarsaDH022,DBLP:conf/onward/PotterMJO22}. This scenario is motivated by an AI-based educational agent acting as a \emph{digital tutor for a student} and providing help by explaining a specific part of a given program that the student is trying to understand. In contrast to the hint generation scenario in Section~\ref{sec.hintgeneration} where the input is a student's buggy program, here the input is a correct program along with a specific part (e.g., a line) that the AI agent is expected to explain to the student. Next, we provide details of this scenario's prompt, input-output formats, performance metrics, and results.

\paragraph{Prompt and output generation.} We begin by describing the content provided as input to a method and the desired output content we seek to generate. The input consists of a detailed \emph{problem description}, a given \emph{program} without bugs, and a \emph{specific part} of the program that a student is trying to understand; the desired output consists of an \emph{explanation} that describes this specific part in the context of the whole program.\footnote{In our evaluation, we obtain these input programs from $25$ buggy programs by fixing all bugs; importantly, programs used as input for this scenario are correct. For a given program, we select a specific part as the program line with the most depth in the Abstract Syntax Tree representation of the program (in the case of ties, we select a line with the higher line number in the program).\label{footnote.contextualized}} Figure~\ref{fig.contextualized_explanation.prompt} shows the prompt---with placeholders for the inputs---used to interact with LLMs for \TechChatGPT{} and \TechGPTFour{} methods.
When interacting with LLMs, we first generate content using this prompt and then manually extract the contextualized explanation as the final output used for evaluation.

\begin{figure}[t!]
    \centering
    \scalebox{0.96}{
        \setlength\tabcolsep{5pt}
        \renewcommand{\arraystretch}{1.2}
\begin{tabular}{||p{0.99\linewidth}||}
    \hline
    \multicolumn{1}{||c||}{\promptheader{Prompt: Contextualized Explanation}} \\ 
    I'm trying to understand a given program for a Python programming problem. Can you help by explaining a specific part of this program? Below I first provide the problem description, then the program, and then a specific part of this program.
    \newline
    \newline
    \promptinput{\{problem\_description\}}
    \newline
    \newline
    Program:
    \newline
    \newline
    \`{}\`{}\`{}\vspace{-2mm}\\
    \promptinput{\{program\}}\\
    \`{}\`{}\`{}
    \newline
    \newline
    Specific Part:
    \newline
    \newline
    \`{}\`{}\`{}\vspace{-2mm}\\
    \promptinput{\{program\_part\_to\_explain\}}\\
    \`{}\`{}\`{}
    \newline
    \newline 
    Can you provide a detailed explanation about the specific part above in the context of the whole program?
    \\
    \hline
\end{tabular}

    }
    \caption{Prompt for the contextualized explanation scenario. This prompt has two placeholders for the problem description and the program, and one placeholder for providing specific part of the program to be explained.
    }
    \label{fig.contextualized_explanation.prompt}
\end{figure}

\begin{figure*}[t!]
\centering
    \begin{subfigure}[b]{1\textwidth}
        \centering
        \scalebox{0.78}{
        \setlength\tabcolsep{10.0pt}
        \renewcommand{\arraystretch}{1.3}
\begin{tabular}{l||cccc}
    \toprule
    \multicolumn{1}{c||}{\textbf{Method}} & \multicolumn{4}{c}{\textbf{Explanation}}\\
     & \multicolumn{1}{c}{Overall} & \multicolumn{1}{c}{Correct} & \multicolumn{1}{c}{Complete} & \multicolumn{1}{c}{Comprehensible} \\  
    \midrule
    \TechChatGPT & $72.0~(12.0)$ & $76.0~(12.0)$ & $100.0~(0.0)$ & $\ \ 94.0~(2.0)$ \\
    \TechGPTFour & $84.0~(\ \ 4.0)$ & $88.0~(\ \ 8.0)$ & $\ \ 98.0~(2.0)$ & $\ \ 96.0~(0.0)$ \\
    {\cellcolor{\tutorcellcolor}\TechTutor} & {\cellcolor{\tutorcellcolor}$92.0~(\ \ 4.0)$} & {\cellcolor{\tutorcellcolor}$92.0~(\ \ 4.0)$} & {\cellcolor{\tutorcellcolor}$100.0~(0.0)$} & {\cellcolor{\tutorcellcolor}$100.0~(0.0)$} \\
    \bottomrule
\end{tabular}

        }
        \vspace{-1mm}
        \caption{}
        \label{fig.contextualized_explanation.results.table}
        \vspace{1mm}
    \end{subfigure}
    \\
    %
    \begin{subfigure}[b]{1\textwidth}
        \centering
        \includegraphics[height=3.8cm]{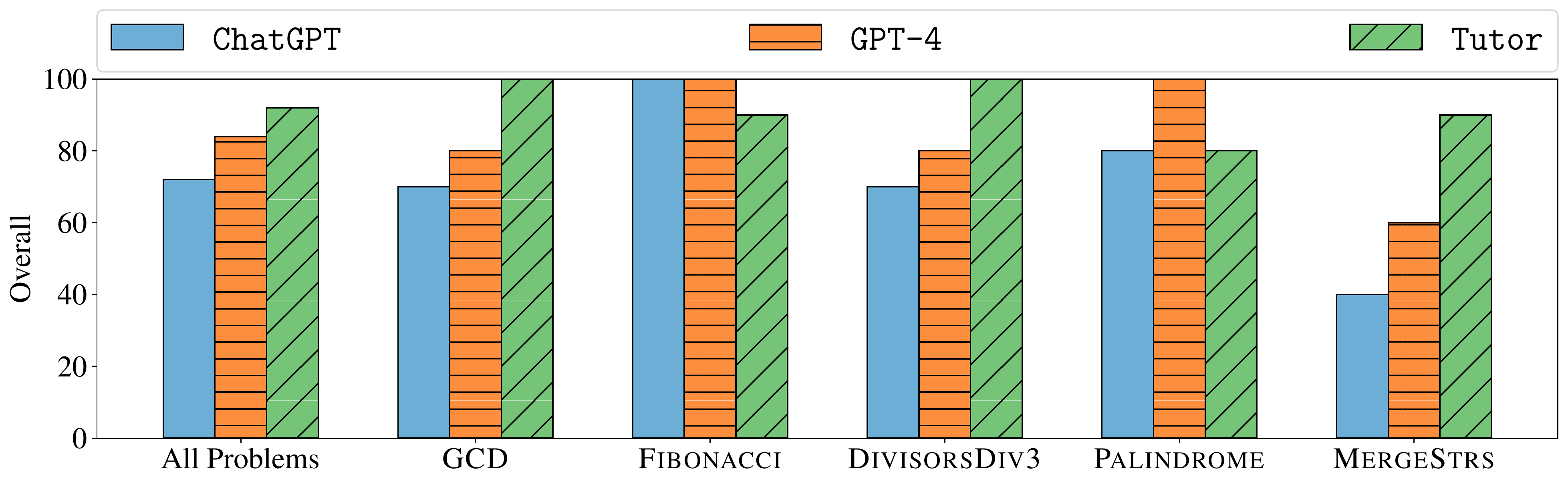}
        \vspace{-2mm}
        \caption{}
        \label{fig.contextualized_explanation.results.histogram}
    \end{subfigure}
    \caption{Results for the contextualized explanation scenario. \textbf{(a)} Results for various metrics aggregated across all problems. \textbf{(b)}~Results for the metric \emph{Overall} separately on five problems. For all metrics, these aggregated results are reported in terms of \%. Details are in Section~\ref{sec.contextualizedexplanation}.}
    \label{fig.contextualized_explanation.results}
    \vspace{-2.5mm}    
\end{figure*}

\paragraph{Output quality and performance metrics.} We assess the generated output along several quality attributes and use aggregated results over these quality attributes as performance metrics in our evaluation.  All attributes for this scenario are binary, with a value of $1$ being better. \emph{Correct} (binary) captures whether the generated explanation contains correct information about the specific part in the context of the whole program. \emph{Complete} (binary) captures whether the generated explanation contains complete information in the context of the whole program. \emph{Comprehensible} (binary) captures whether the generated explanation is easy to understand, presented in a readable format, and doesn't contain redundant information. \emph{Overall} (binary) is $1$ only when if the generated explanation satisfies \emph{all} the three attributes mentioned above. Human evaluators annotate the quality of generated output for each of the $25$ instances; in this scenario, all the above attributes require manual annotation (similar to the hint generation scenario in Section~\ref{sec.hintgeneration}).

\paragraph{Results.}
Figure~\ref{fig.contextualized_explanation.results.table} provide results for various metrics aggregated across all problems, and Figure~\ref{fig.contextualized_explanation.results.histogram} for the metric \emph{Overall} separately on five problems. These aggregated results for various metrics are reported in terms of \%. Next, we summarize some of the key findings. First, results in Figure~\ref{fig.contextualized_explanation.results.table} for the metric \emph{Overall} highlight that \TechGPTFour{} ($84.0$) and \TechChatGPT{} ($72.0$) have high performance, close to that of the performance of \TechTutor{} ($92.0$). One of the main reasons for this high performance in this scenario is that the methods take bug-free programs as input; moreover, the input programs are short and correspond to solutions to popular problems, which makes it somewhat easy to provide correct contextualized explanations. Second, results in Figure~\ref{fig.contextualized_explanation.results.histogram} highlight that these findings are generally consistent across all five problems for the metric \emph{Overall}; the gap in the performance of \TechGPTFour{} vs. \TechTutor{} is worst on \PMergeStrings{} for this scenario. In Appendix~\ref{app.sec.contextualizedexplanation}, we provide an illustrative example to qualitatively show the outputs of different methods.

\section{Task Synthesis Scenario}\label{sec.taskcreation}
This section is dedicated to the programming education scenario of \emph{task synthesis}~\cite{DBLP:conf/nips/AhmedCEFGRS20,DBLP:conf/aied/GhoshTDS22,DBLP:conf/icer/SarsaDH022,neurtasksyn}. This scenario is motivated by an AI-based educational agent acting as a \emph{digital assistant for an educator} or \emph{digital tutor for a student} -- the agent provides assistance/help by generating new tasks (in the form of debugging quizzes) that exercise specific types of bugs the student is encountering. Next, we provide details of this scenario's prompt, input-output formats, performance metrics, and results.

\looseness-1\paragraph{Prompt and output generation.} We begin by describing the content provided as input to a method and the desired output content we seek to generate. The input consists of a detailed \emph{problem description}, a student's \emph{buggy program}, and fixes to the buggy program as \emph{line-diffs with fixed program}; the desired output consists of a new debugging task comprising \emph{(new problem, new buggy program)}.\footnote{In our evaluation, we obtain these fixed programs from $25$ buggy programs by fixing all bugs with a minimal number of edits required.\label{footnote.taskcreation}} Figure~\ref{fig.task_creation.prompt} shows the prompt used to interact with LLMs for \TechChatGPT{} and \TechGPTFour{} methods. When interacting with LLMs, we first generate content using this prompt and then manually extract the new debugging task, i.e., (new problem, new buggy program) as the final output for evaluation.
%

\begin{figure}[t!]
    \centering
    \scalebox{0.96}{
        \setlength\tabcolsep{5pt}
        \renewcommand{\arraystretch}{1.2}
\begin{tabular}{||p{0.99\linewidth}||}
    \hline
    \multicolumn{1}{||c||}{\promptheader{Prompt: Task Synthesis}} \\ 
    I'm helping a novice student on a Python programming problem. The student's program below has bug(s). Can you help by creating a new simpler problem along with a minimal buggy program that highlights a bug in the student’s program? Below I first provide the problem description, the student’s buggy program, and then a fix to the student’s buggy program.
    \newline
    \newline
    \promptinput{\{problem\_description\}}
    \newline
    \newline
    Student's Buggy Program:
    \newline
    \newline
    \`{}\`{}\`{}\vspace{-2mm}\\
    \promptinput{\{buggy\_program\}}\\
    \`{}\`{}\`{}
    \newline
    \newline
    Fix to the Student's Buggy Program:
    \newline
    \newline
    \`{}\`{}\`{}\vspace{-2mm}\\
    \promptinput{\{line\_diffs\_with\_fixed\_program\}}\\
    \`{}\`{}\`{}
    \newline
    \newline 
    Based on the above, can you create a new simpler problem along with a buggy program that has the same type of bug(s)? If the student’s buggy program has multiple bugs, it is ok to focus on only one of those bugs. Make sure that the new problem is simpler than the original problem.
    \\
    \hline
\end{tabular}

    }
    \caption{Prompt for the task synthesis scenario. This prompt has two placeholders for the problem description and the student's buggy program, and one placeholder for providing fix to the student's buggy program in the form of line differences.
    }
    \label{fig.task_creation.prompt}
    \vspace{-2.5mm}
\end{figure}

\begin{figure*}[t!]
\centering
    \begin{subfigure}[b]{1\textwidth}
        \centering
        \scalebox{0.78}{
        \setlength\tabcolsep{3pt}
        \renewcommand{\arraystretch}{1.3}
\begin{tabular}{l||c|cc|cc}
    \toprule
    \multicolumn{1}{c||}{\textbf{Method}} & \multicolumn{1}{c|}{\textbf{(New Problem, New Buggy Program)}} & \multicolumn{2}{c|}{\textbf{New Problem}} & \multicolumn{2}{c}{\textbf{New Buggy Program}}\\
     & \multicolumn{1}{c|}{Overall} & \multicolumn{1}{c}{Correct} & \multicolumn{1}{c|}{Simpler} & \multicolumn{1}{c}{SimilarBugs} & \multicolumn{1}{c}{MinimalBugs} \\  
    \midrule
    \TechChatGPT & $10.0~(2.0)$ & $78.0~(10.0)$ & $66.0~(2.0)$ & $36.0~(0.0)$ & $76.0~(8.0)$ \\
    \TechGPTFour & $22.0~(2.0)$ & $94.0~(\ \ 6.0)$ & $88.0~(4.0)$ & $40.0~(0.0)$ & $76.0~(8.0)$ \\
    {\cellcolor{\tutorcellcolor}\TechTutor} & {\cellcolor{\tutorcellcolor}$74.0~(2.0)$} & {\cellcolor{\tutorcellcolor}$98.0~(\ \ 2.0)$} & {\cellcolor{\tutorcellcolor}$98.0~(2.0)$} & {\cellcolor{\tutorcellcolor}$92.0~(4.0)$} & {\cellcolor{\tutorcellcolor}$82.0~(6.0)$} \\
    \bottomrule
\end{tabular}
        }
        \vspace{-1mm}
        \caption{}
        \label{fig.task_creation.results.table}
        \vspace{1mm}
    \end{subfigure}
    \\
    %
    \begin{subfigure}[b]{1\textwidth}
        \centering
        \includegraphics[height=3.8cm]{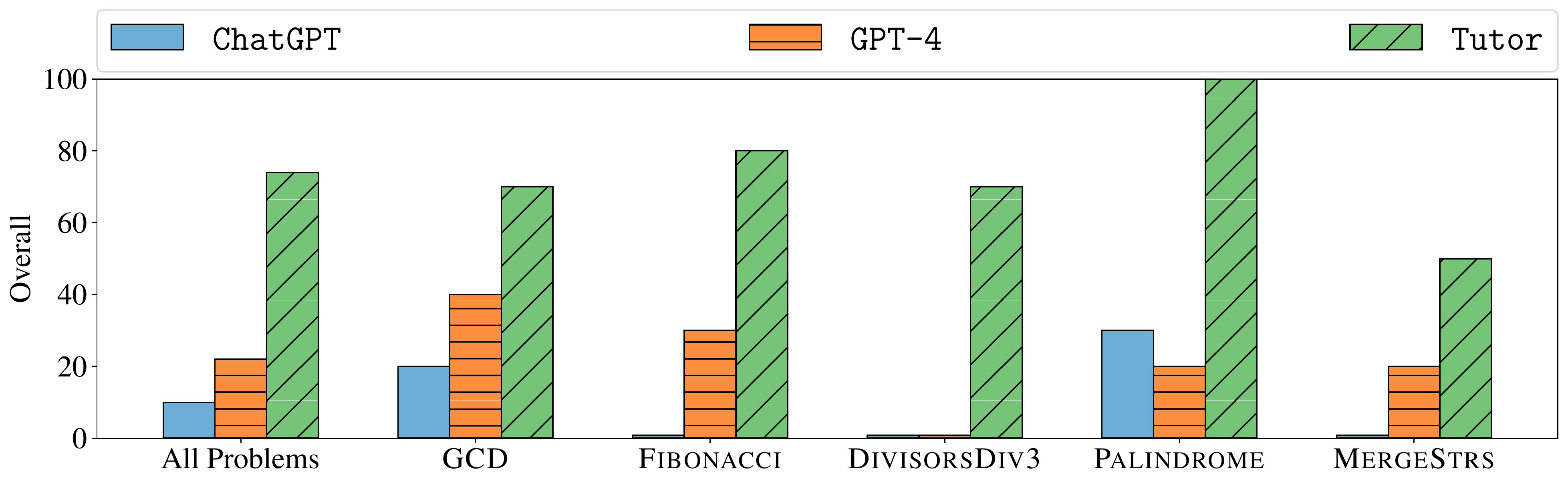}
        \vspace{-2mm}
        \caption{}
        \label{fig.task_creation.results.histogram}
    \end{subfigure}
    \caption{Results for the task synthesis scenario. \textbf{(a)} Results for various metrics aggregated across all problems. \textbf{(b)} Results for the metric \emph{Overall} separately on five problems. For all metrics, these aggregated results are reported in terms of \%. Details are in Section~\ref{sec.taskcreation}.}
    \label{fig.task_creation.results}
    \vspace{-2.5mm}    
\end{figure*}

\paragraph{Output quality and performance metrics.} We assess the generated output along several quality attributes and use aggregated results over these quality attributes as performance metrics in our evaluation.  All attributes for this scenario are binary, with a value of $1$ being better. \emph{Correct} (binary) captures whether the generated new problem is correct in terms of its description and specification, and can be solved. \emph{Simpler} (binary) captures whether the generated new problem is simpler than the input problem. \emph{SimilarBugs} (binary) captures whether the generated new buggy program has bug(s) similar to bug(s) in the student's buggy program. \emph{MinimalBugs} (binary) captures that the generated new buggy program doesn't contain any other bugs. \emph{Overall} (binary) is $1$ only if the generated new problem and new buggy program jointly satisfy \emph{all} the four attributes mentioned earlier. Human evaluators annotate the quality of generated output for each of the $25$ instances; in this scenario, all the above attributes require manual annotation.

\looseness-1\paragraph{Results.} Figure~\ref{fig.task_creation.results.table} provide results for various metrics aggregated across all problems, and Figure~\ref{fig.task_creation.results.histogram} for the metric \emph{Overall} separately on five problems. These aggregated results for various metrics are
reported in terms of \%. Next, we summarize some of the key findings. First, results in Figure~\ref{fig.task_creation.results.table} for the metric \emph{Overall} highlight that even though \TechGPTFour{} ($22.0$) improves up on \TechChatGPT{} ($10.0$), it still performs substantially worse in comparison to the performance of \TechTutor{} ($74.0$). Further, results on various metrics suggest that the gap in the performance of \TechGPTFour{} and \TechTutor{} is worse on the metric \emph{SimilarBugs} that requires a more in-depth understanding of bugs in the input program and then transferring them to a newly generated program. Second, results in Figure~\ref{fig.task_creation.results.histogram} highlight that these findings are generally consistent across all five problems for the metric \emph{Overall}; the gap in the performance of \TechGPTFour{} vs. \TechTutor{} is worst on \PDivisors{} and \PPalindrome{}. In Appendix~\ref{app.sec.taskcreation}, we provide an illustrative example to qualitatively show the outputs of different methods.
%

\section{Concluding Discussions}\label{sec.conclusion}
We conducted a study to benchmark state-of-the-art generative AI and large language models for a comprehensive set of programming education scenarios. Our results show that GPT-4 drastically outperforms ChatGPT (based on GPT-3.5) and comes close to human tutors' performance for several scenarios. These results also highlight scenarios and specific problems where GPT-4 still struggles, in particular, for the scenarios of grading feedback and task synthesis that have a substantial gap in the performance of GPT-4 compared to that of human tutors.

\looseness-1Next, we discuss some limitations of our current work and ideas to tackle them in the future. First, our work involved only two human experts acting as tutors and evaluators; it would be useful to scale up the study. Second, we focused only on introductory Python programming education; it would be interesting to conduct a similar study for other programming languages and other domains beyond programming. Third, we considered English as the primary mode of language, and it would be interesting to evaluate these models in multilingual settings. Fourth, our evaluation only considered expert-based assessments and didn't involve students; it would be useful to consider student-based assessments.

Apart from the above extensions, there are many exciting directions for future work, including but not limited to: (a) curating larger-scale benchmarks that the research community can use to evaluate new versions of these models; (b) evaluating alternate generative models, in particular, open-source variants; (c) developing techniques to improve the performance of generative AI and large language models, e.g., by leveraging symbolic methods, fine-tuning, or automated prompting; (d) conducting studies in classrooms with students.

%
\begin{ack}
Funded/Co-funded by the European Union (ERC, TOPS, 101039090). Views and opinions expressed are however those of the author(s) only and do not necessarily reflect those of the European Union or the European Research Council. Neither the European Union nor the granting authority can be held responsible for them.
\end{ack}

\bibliographystyle{unsrtnat}
\bibliography{main}

\clearpage
\begin{appendices}

%
\section{Appendix}

This appendix provides illustrative examples for six programming education scenarios discussed in Sections~\ref{sec.programrepair}--\ref{sec.taskcreation}. For each scenario, we have picked one illustrative example to highlight settings where GPT-4 still struggles. These examples provide further insights and potential ideas for future work on developing techniques to improve the performance of these models. For each example, we first show specific inputs provided in the prompt, followed by the outputs from \TechChatGPT, \TechGPTFour, and \TechTutor. The detailed problem descriptions used in prompts are mentioned in Footnote~\ref{footnote.problemdescription}.

When presenting these illustrative examples, we obfuscate the input programs by changing variable names and formatting styles while keeping the original bugs exactly the same. Accordingly, we make the same modifications in the generated output to align it with these changes. 

Next, we provide a brief description of the content in the rest of this appendix.
\begin{itemize}[leftmargin=*]
\item \looseness-1Appendix~\ref{app.sec.problemsetup} provides solution programs generated by \TechGPTFour{} for all problems.
\item Appendix~\ref{app.sec.programrepair} provides an example for the program repair scenario on \PFibonacci{}.
\item Appendix~\ref{app.sec.hintgeneration} provides an example for the hint generation scenario on \PPalindrome{}.
\item Appendix~\ref{app.sec.gradingfeedback} provides an example for the grading feedback scenario on \PFibonacci{}.
\item Appendix~\ref{app.sec.pairprogramming} provides an example for the pair programming scenario on \PDivisors{}.
\item Appendix~\ref{app.sec.contextualizedexplanation} provides an example for the contextualized explanation scenario on \PMergeStrings{}.
\item Appendix~\ref{app.sec.taskcreation} provides an example for the task synthesis scenario on \PGCD{}.
\end{itemize}

\subsection{Solution Programs for Five Problems}
\label{app.sec.problemsetup}

Figure~\ref{fig.problemsetup.solutions.Fibonacci} in Section~\ref{sec.problemsetup} provided a solution program generated by  \TechGPTFour{} for \PFibonacci{}. Figures~\ref{fig.problemsetup.solutions.MergeStrings},~\ref{fig.problemsetup.solutions.GCD},~\ref{fig.problemsetup.solutions.Divisors},~and~\ref{fig.problemsetup.solutions.Palindrome} below provide solution programs generated by  \TechGPTFour{} for the remaining four problems. \TechGPTFour{} is able to correctly solve these five problems, i.e., these generated programs pass automated test suites (see~Footnote~\ref{footnote.problemdescription}).

\begin{figure}[h!]
    \centering
    \begin{subfigure}[b]{1.03\textwidth}
        \centering
        \scalebox{0.9}{
            \renewcommand{\arraystretch}{1}
            \begin{tabular}{|p{1\linewidth}|}
                \hline
                \multicolumn{1}{|p{1\linewidth}|}{
\begin{lstlisting}[basicstyle=\fontsize{7}{7}\ttfamily,xleftmargin=1.65em,belowskip=-0.7em,aboveskip=-0.2em,
                    linebackgroundcolor={
                    \ifnum\value{lstnumber}>3
                        \ifnum\value{lstnumber}<21
                            \color{CodeHighlight}
                        \fi
                    \fi
                    }]
#User function Template for python3
class Solution:
    def merge(self, S1, S2):
        # initializing the resultant string
        result = ""
        # getting the length of both strings
        len1, len2 = len(S1), len(S2)
        # getting the minimum length
        min_len = min(len1, len2)
        # adding characters alternatively from both strings to the resultant string
        for i in range(min_len):
            result += S1[i] + S2[i]
        # if S1 is longer, adding the remaining characters of S1 to the resultant string
        if len1 > len2:
            result += S1[min_len:]
        # if S2 is longer, adding the remaining characters of S2 to the resultant string
        elif len1 < len2:
            result += S2[min_len:]
        # returning the resultant string
        return result

#{ 
# Driver Code Starts
#Initial Template for Python 3
if __name__ == '__main__': 
    t = int(input())
    for _ in range(t):
        S1,S2 = map(str,input().strip().split())
        ob = Solution()
        print(ob.merge(S1, S2))
# } Driver Code Ends
\end{lstlisting}}\\
                \hline
            \end{tabular}
        }
    \end{subfigure}    
    \caption{A solution program generated by  \TechGPTFour{} for \PMergeStrings{}.}
    \vspace{5mm}
    \label{fig.problemsetup.solutions.MergeStrings}
\end{figure}
\begin{figure}[h!]
    \centering
    \begin{subfigure}[b]{1.03\textwidth}
        \centering
        \scalebox{0.9}{
            \renewcommand{\arraystretch}{1}
            \begin{tabular}{|p{1\linewidth}|}
                \hline
                \multicolumn{1}{|p{1\linewidth}|}{
\begin{lstlisting}[basicstyle=\fontsize{7}{7}\ttfamily,xleftmargin=1.65em,belowskip=-0.7em,aboveskip=-0.2em,
                    linebackgroundcolor={
                    \ifnum\value{lstnumber}>3
                        \ifnum\value{lstnumber}<11
                            \color{CodeHighlight}
                        \fi
                    \fi
                    }]
#User function Template for python3
class Solution:
    def gcd(self, A, B):
        # If B is 0 then A is the GCD
        if B == 0:
            return A
        # Else we recursively call the gcd function
        # with B and the remainder when A is divided by B
        else:
            return self.gcd(B, A % B)

# { 
# Driver Code Starts
#Initial Template for Python 3

if __name__ == '__main__': 
    t = int(input())
    for _ in range(t):
        A,B = list(map(int, input().strip().split()))
        ob = Solution()
        print(ob.gcd(A,B))
# } Driver Code Ends
\end{lstlisting}}\\
                \hline
            \end{tabular}
        }
    \end{subfigure}
    \caption{A solution program generated by  \TechGPTFour{} for \PGCD{}.}
    \vspace{5mm}
    \label{fig.problemsetup.solutions.GCD}
\end{figure}
\begin{figure}[h!]
    \centering
    \begin{subfigure}[b]{1.03\textwidth}
        \centering
        \scalebox{0.9}{
            \renewcommand{\arraystretch}{1}
            \begin{tabular}{|p{1\linewidth}|}
                \hline
                \multicolumn{1}{|p{1\linewidth}|}{
\begin{lstlisting}[basicstyle=\fontsize{7}{7}\ttfamily,xleftmargin=1.65em,belowskip=-0.7em,aboveskip=-0.2em,
                    linebackgroundcolor={
                    \ifnum\value{lstnumber}>5
                        \ifnum\value{lstnumber}<17
                            \color{CodeHighlight}
                        \fi
                    % \fi
                    \else
                        \ifnum\value{lstnumber}=2
                            \color{CodeHighlight}
                        \fi
                    \fi
                    }]
#User function Template for python3
import math

class Solution:
    def count_divisors(self, N):
        count = 0
        i = 1
        while i * i <= N:
            if N % i == 0:
                # Check if divisors are divisible by 3
                if i % 3 == 0:
                    count += 1
                if i != N // i and (N // i) % 3 == 0:
                    count += 1
            i += 1
        return count

# Driver Code Starts
#Initial Template for Python 3
if __name__ == '__main__': 
    t = int (input ())
    for _ in range (t):
        N = int(input())
        ob = Solution()
        print(ob.count_divisors(N))
# } Driver Code Ends
\end{lstlisting}}\\
                \hline
            \end{tabular}
        }
    \end{subfigure}
    \caption{A solution program generated by  \TechGPTFour{} for \PDivisors{}.}
    \vspace{5mm}
    \label{fig.problemsetup.solutions.Divisors}
\end{figure}

\begin{figure}[h!]
    \centering
    \begin{subfigure}[b]{1.03\textwidth}
        \centering
        \scalebox{0.9}{
            \renewcommand{\arraystretch}{1}
            \begin{tabular}{|p{1\linewidth}|}
                \hline
                \multicolumn{1}{|p{1\linewidth}|}{
\begin{lstlisting}[basicstyle=\fontsize{7}{7}\ttfamily,xleftmargin=1.65em,belowskip=-0.7em,aboveskip=-0.2em,
                    linebackgroundcolor={
                    \ifnum\value{lstnumber}>3
                        \ifnum\value{lstnumber}<13
                            \color{CodeHighlight}
                        \fi
                    \fi
                    }]
#User function Template for python3
class Solution:
    def isPalindrome(self, S):
        # Start from the ends and work inwards
        left, right = 0, len(S) - 1
        while left < right:
            # If the characters don't match, it's not a palindrome
            if S[left] != S[right]:
                return 0
            left, right = left + 1, right - 1
        # All characters matched, it's a palindrome
        return 1

#{
# Driver Code Starts
#Initial Template for Python 3
if __name__ == '__main__':
    T=int(input())
    for i in range(T):
        S = input()
        ob = Solution()
        answer = ob.isPalindrome(S)
        print(answer)
# } Driver Code Ends
\end{lstlisting}}\\
                \hline
            \end{tabular}
        }
    \end{subfigure}
    \caption{A solution program generated by  \TechGPTFour{} for \PPalindrome{}.}
    \vspace{5mm}
    \label{fig.problemsetup.solutions.Palindrome}
\end{figure}

\clearpage

\subsection{Illustrative Example for the Program Repair Scenario}
\label{app.sec.programrepair}

Figure~\ref{fig.program_repair.example} shows an example for the program repair scenario on \PFibonacci{} problem. In this example, the student's buggy program shown in Figure~\ref{fig.program_repair.example.input} misses some numbers at the end of the series for a few test cases. In the generated programs shown in Figures~\ref{fig.program_repair.example.output_ChatGPT},~\ref{fig.program_repair.example.output_GPT4},~and~\ref{fig.program_repair.example.output_Tutor}, we have highlighted the lines where the input buggy program has been modified. \TechChatGPT{} generates an incorrect program (i.e., the  attribute \emph{Correct} is $0$). \TechGPTFour{} generates a correct program (i.e., the attribute \emph{Correct} is $1$); however, it makes large number of edits (\emph{EditTokens} is $33$). \TechTutor{} generates a  correct program with a small number of edits (i.e., \emph{EditTokens} is $1$).

%

\begin{figure}[h!]
    \centering
    \begin{subfigure}[b]{0.77\textwidth}
        \centering
        \scalebox{0.85}{
            \renewcommand{\arraystretch}{1}
            \begin{tabular}{|p{1\linewidth}|}
                \hline
                \multicolumn{1}{|p{1\linewidth}|}{
\begin{lstlisting}[basicstyle=\fontsize{8.5}{8.5}\ttfamily,xleftmargin=1.65em,belowskip=-0.7em,aboveskip=-0.2em]
class Solution:
    def nFibonacci(self, N):
        res = [0, 1]
        for i in range(2, N + 1):
            if(res[i - 1] + res[i - 2] <= N):
                res.append(res[i - 1] + res[i - 2])
            else:
                break
        return(res)
\end{lstlisting}}\\
                \hline
            \end{tabular}
        }
        \caption{Input: Buggy program}				
        \vspace{5mm}
        \label{fig.program_repair.example.input}
    \end{subfigure}
    \begin{subfigure}[b]{0.77\textwidth}
        \centering
        \scalebox{0.85}{
            \renewcommand{\arraystretch}{1}
            \begin{tabular}{|p{1\linewidth}|}
                \hline
                \multicolumn{1}{|p{1\linewidth}|}{
\begin{lstlisting}[basicstyle=\fontsize{8.5}{8.5}\ttfamily,xleftmargin=1.65em,belowskip=-0.7em,aboveskip=-0.2em,linebackgroundcolor={
            \ifnum\value{lstnumber}=4\color{CodeHighlight}\fi
        }]
class Solution:
    def nFibonacci(self, N):
        res = [0, 1]
        for i in range(2, N):
            if res[i - 1] + res[i - 2] <= N:
                res.append(res[i - 1] + res[i - 2])
            else:
                break
        return res
\end{lstlisting}}\\
                \hline
            \end{tabular}
        }
        \subcaption{Output by \TechChatGPT}		\vspace{5mm}
        \label{fig.program_repair.example.output_ChatGPT}
    \end{subfigure}
    \begin{subfigure}[b]{0.77\textwidth}
        \centering
        \scalebox{0.85}{
            \renewcommand{\arraystretch}{1}
            \begin{tabular}{|p{1\linewidth}|}
                \hline
                \multicolumn{1}{|p{1\linewidth}|}{
\begin{lstlisting}[basicstyle=\fontsize{8.5}{8.5}\ttfamily,xleftmargin=1.65em,belowskip=-0.7em,aboveskip=-0.2em,linebackgroundcolor={
            \ifnum\value{lstnumber}=4\color{CodeHighlight}\fi
            \ifnum\value{lstnumber}=5\color{CodeHighlight}\fi
            \ifnum\value{lstnumber}=6\color{CodeHighlight}\fi
            \ifnum\value{lstnumber}=7\color{CodeHighlight}\fi
        }]
class Solution:
    def nFibonacci(self, N):
        res = [0, 1]
        while True:
            next_val = res[-1] + res[-2]
            if next_val <= N:
                res.append(next_val)
            else:
                break
        return(res)
\end{lstlisting}}\\
                \hline
            \end{tabular}
        }
        \subcaption{Output by \TechGPTFour}		\vspace{5mm}
        \label{fig.program_repair.example.output_GPT4}
    \end{subfigure}
    \begin{subfigure}[b]{0.77\textwidth}
        \centering
        \scalebox{0.85}{
            \renewcommand{\arraystretch}{1}
            \begin{tabular}{|p{1\linewidth}|}
                \hline
                \multicolumn{1}{|p{1\linewidth}|}{
\begin{lstlisting}[basicstyle=\fontsize{8.5}{8.5}\ttfamily,xleftmargin=1.65em,belowskip=-0.7em,aboveskip=-0.2em,linebackgroundcolor={
            \ifnum\value{lstnumber}=4\color{CodeHighlight}\fi
        }]
class Solution:
    def nFibonacci(self, N):
        res = [0, 1]
        for i in range(2, N + 2):
            if res[i - 1] + res[i - 2] <= N:
                res.append(res[i - 1] + res[i - 2])
            else:
                break
        return(res)
\end{lstlisting}}\\
                \hline
            \end{tabular}
        }
        \subcaption{Output by \TechTutor}
        \vspace{5mm}
        \label{fig.program_repair.example.output_Tutor}
    \end{subfigure}
    \caption{An illustrative example for the program repair scenario on \PFibonacci{} problem.}
	\label{fig.program_repair.example}
\end{figure}

\clearpage

\subsection{Illustrative Example for Hint Generation Scenario}
\label{app.sec.hintgeneration}

\looseness-1Figure~\ref{fig.hint_generation.example} shows an example for the hint generation scenario on \PPalindrome{} problem. In this example, the student's buggy program shown in Figure~\ref{fig.hint_generation.example.input} wrongly considers all strings with odd lengths as not palindrome. In the generated outputs shown in Figures~\ref{fig.hint_generation.example.output_ChatGPT},~\ref{fig.hint_generation.example.output_GPT4},~and~\ref{fig.hint_generation.example.output_Tutor}, we have highlighted the text where the hint/explanation is wrong. Both \TechChatGPT{} and \TechGPTFour{} generate an incorrect hint (i.e., the attribute \emph{HintCorrect} is $0$). \TechTutor{}'s output meets all the quality criteria  (i.e., \emph{Overall} is $1$).

%

\begin{figure}[h!]
    \centering
    \begin{subfigure}[b]{0.5\textwidth}
        \centering
        \scalebox{0.85}{
            \renewcommand{\arraystretch}{1}
            \begin{tabular}{|p{1\linewidth}|}
                \hline
                \multicolumn{1}{|p{1\linewidth}|}{
\begin{lstlisting}[basicstyle=\fontsize{8.5}{7}\ttfamily,xleftmargin=1.65em,belowskip=-0.7em,aboveskip=-0.2em]
class Solution:
    def isPalindrome(self, S):
        if len(S) % 2 != 0:
            return 0
        for i in range(len(S)):
            if S[i] != S[-i - 1]:
                return 0
        return 1
\end{lstlisting}}\\
                \hline
            \end{tabular}
        }
        \subcaption{Input: Buggy program}
        \vspace{5mm}
        \label{fig.hint_generation.example.input}
    \end{subfigure}
    \allowbreak
    
    \begin{subfigure}[t!]{0.97\textwidth}
    {
        \centering
        \scalebox{0.9}{
            \renewcommand{\arraystretch}{1.6}
            \begin{tabular}{|p{1\linewidth}|}
                \hline
                \multicolumn{1}{|c|}{\componentheader{Hint}}\\
                \multicolumn{1}{|p{1\linewidth}|}{
                      \input{figs/4_hint_generation/fig_examples_data_new/Q8_p2_ChatGPT}
                }\\
                \hline
                \multicolumn{1}{|c|}{\componentheader{Explanation}}\\
                % ...
                \multicolumn{1}{|p{1\linewidth}|}{
                [...] The program has two bugs: \newline \textcolor{ExpHighlight}{- The condition `if len(S) \% 2 != 0' incorrectly checks if the length of the string is odd, but it should check if it is even.} \newline - The loop `for i in range(len(S))' iterates through [...]
                }\\
                % ...
                \hline
            \end{tabular}
        }
        \vspace{-1mm}
        \caption{Output by \TechChatGPT}		
        \vspace{5mm}
        \label{fig.hint_generation.example.output_ChatGPT}
    }
    \end{subfigure}
    
    \begin{subfigure}[t!]{0.97\textwidth}
    {
        \centering
        \scalebox{0.9}{
            \renewcommand{\arraystretch}{1.6}
            \begin{tabular}{|p{1\linewidth}|}
                \hline
                \multicolumn{1}{|c|}{\componentheader{Hint}}\\
                \multicolumn{1}{|p{1\linewidth}|}{
                      \input{figs/4_hint_generation/fig_examples_data_new/Q8_p2_GPT4}
                }\\
                \hline
                \multicolumn{1}{|c|}{\componentheader{Explanation}}\\
                % ...
                \multicolumn{1}{|p{1\linewidth}|}{
                There are two main bugs in your code: \newline - First, your assumption that the string cannot be a palindrome if its length is odd is incorrect. [...] \newline \textcolor{ExpHighlight}{- Secondly, you are comparing characters at mirrored indices incorrectly.} Python allows for negative indexing, with `-1' being the last element, `-2' the second last, and so forth. In your code, you're comparing `S[I]' with `S[-i-1]', which will actually compare the first character with the last, second with second last and so on, \textcolor{ExpHighlight}{but when `i' reaches the midpoint of the string, `-i-1' becomes positive and starts from the start of the string, which is incorrect.}
                }\\
                % ...
                \hline
            \end{tabular}
        }
        \vspace{-1mm}
        \caption{Output by \TechGPTFour}		
        \vspace{5mm}
        \label{fig.hint_generation.example.output_GPT4}
    }
    \end{subfigure}
    
    \begin{subfigure}[t!]{0.97\textwidth}
    {
        \centering
        \scalebox{0.9}{
            \renewcommand{\arraystretch}{1.6}
            \begin{tabular}{|p{1\linewidth}|}
                \hline
                \multicolumn{1}{|c|}{\componentheader{Hint}}\\
                \multicolumn{1}{|p{1\linewidth}|}{
                      \input{figs/4_hint_generation/fig_examples_data_new/Q8_p2_Tutor}
                }\\
                \hline
                \multicolumn{1}{|c|}{\componentheader{Explanation}}\\
                
                \multicolumn{1}{|p{1\linewidth}|}{
                For Input: pjxcxjp
                \newline
                Your Code's output is: 0
                \newline
                It's Correct output is: 1
                \newline
                Look at line 3-4. Note that a string with an odd number of characters can also be palindrome.
                }\\
                \hline
            \end{tabular}
        }
        \vspace{-1mm}
        \caption{Output by \TechTutor}		
        \vspace{5mm}
        \label{fig.hint_generation.example.output_Tutor}
    }
    \end{subfigure}

    \caption{An illustrative example for the hint generation scenario on \PPalindrome{} problem.}
	\label{fig.hint_generation.example}
\end{figure}

\clearpage

\subsection{Illustrative Example for Grading Feedback Scenario}
\label{app.sec.gradingfeedback}

Figure~\ref{fig.grading_feedback.example} shows an example for the grading feedback scenario on \PFibonacci{} problem. In this example, the student's buggy program shown in Figure~\ref{fig.grading_feedback.example.input} misses some numbers at the beginning of the series for most test cases. Both \TechChatGPT{} and \TechChatGPT{} fail to recognize this issue in the student's program and wrongly assign high points for the rubric dimensions of \emph{CorrectGeneral} (i.e., Correctness for general inputs) and \emph{CorrectEdge} (i.e., Correctness for edge cases). Interestingly, \TechGPTFour{} also writes a detailed textual description where it \emph{wrongly reasons about} its grading strategy: 

\begin{quote}
\looseness-1``[...] Correctness for general inputs (30 points): The program seems to handle general inputs well, but it does have an issue. The program does not return the correct result for N = 0. However, as per the problem statement, N is always greater than or equal to 1. So, this doesn't impact the grading in this category. Score: 30/30 [...]''.
\end{quote}

\TechTutor{} recognizes this issue in the program and correctly assigns $0$ points for these two dimensions. 
%

%

%

\begin{figure}[h!]
    \centering
    \begin{subfigure}[b]{0.45\textwidth}
        \centering
        \scalebox{0.85}{
            \renewcommand{\arraystretch}{1}
            \begin{tabular}{|p{1\linewidth}|}
                \hline
                \multicolumn{1}{|p{1\linewidth}|}{
\begin{lstlisting}[basicstyle=\fontsize{8.5}{7}\ttfamily,xleftmargin=1.65em,belowskip=-0.7em,aboveskip=-0.2em]
class Solution:
    def nFibonacci(self, N):
        res = []
        if N == 0:
            res.append(0)
        if N == 1:
            res.append(0)
            res.append(1)
            res.append(1)
        x = 1
        y = 1
        z = x + y
        while z <= N:
            res.append(z)
            x, y = y, z
            z = x + y
        return res
\end{lstlisting}}\\
                \hline
            \end{tabular}
        }
        \caption{Input: Buggy program}
        \vspace{5mm}
        \label{fig.grading_feedback.example.input}
    \end{subfigure}
    \allowbreak

    \begin{minipage}{1\linewidth}
        \centering
        \scalebox{0.78}{
            \begin{tabular}{l||cccccc}
                \toprule
                \multicolumn{1}{c||}{\textbf{Method}} & \multicolumn{6}{c}{\textbf{Points Across Grading Rubric}}\\
                 & \multicolumn{1}{c}{Total} & \multicolumn{1}{c}{ProgramFormat} & \multicolumn{1}{c}{TimeComplexity} & \multicolumn{1}{c}{SpaceComplexity} & \multicolumn{1}{c}{CorrectGeneral} & \multicolumn{1}{c}{CorrectEdge} \\  
                \midrule
                \TechChatGPT & $90$ & $10$ & $15$ & $15$ & $30$ & $20$ \\
                \bottomrule
            \end{tabular}
        }
        \subcaption{Output by \TechChatGPT}
        \vspace{5mm}
        \label{fig.grading_feedback.example.output_ChatGPT}
    \end{minipage}

    \begin{minipage}{1\linewidth}
        \centering
        \scalebox{0.78}{
            \begin{tabular}{l||cccccc}
                \toprule
                \multicolumn{1}{c||}{\textbf{Method}} & \multicolumn{6}{c}{\textbf{Points Across Grading Rubric}}\\
                 & \multicolumn{1}{c}{Total} & \multicolumn{1}{c}{ProgramFormat} & \multicolumn{1}{c}{TimeComplexity} & \multicolumn{1}{c}{SpaceComplexity} & \multicolumn{1}{c}{CorrectGeneral} & \multicolumn{1}{c}{CorrectEdge} \\  
                \midrule
                \TechGPTFour & $90$ & $10$ & $15$ & $15$ & $30$ & $20$ \\
                \bottomrule
            \end{tabular}
        }
        \subcaption{Output by \TechGPTFour}
        \vspace{5mm}
        \label{fig.grading_feedback.example.output_GPT4}
    \end{minipage}

    \begin{minipage}{1\linewidth}
        \centering
        \scalebox{0.78}{
            \begin{tabular}{l||cccccc}
                \toprule
                \multicolumn{1}{c||}{\textbf{Method}} & \multicolumn{6}{c}{\textbf{Points Across Grading Rubric}}\\
                 & \multicolumn{1}{c}{Total} & \multicolumn{1}{c}{ProgramFormat} & \multicolumn{1}{c}{TimeComplexity} & \multicolumn{1}{c}{SpaceComplexity} & \multicolumn{1}{c}{CorrectGeneral} & \multicolumn{1}{c}{CorrectEdge} \\  
                \midrule
                \TechTutor & $40$ & $10$ & $15$ & $15$ & $0$ & $0$ \\
                \bottomrule
            \end{tabular}
        }
        \subcaption{Output by \TechTutor}
        \vspace{5mm}
        \label{fig.grading_feedback.example.output_Tutor}
    \end{minipage}
    
    \caption{An illustrative example for the grading feedback scenario on \PFibonacci{} problem.}
	\label{fig.grading_feedback.example}
\end{figure}

\clearpage

\subsection{Illustrative Example for Pair Programming Scenario}
\label{app.sec.pairprogramming}

Figure~\ref{fig.pair_programming.example} shows an example for the pair programming scenario on \PDivisors{} problem. In this example, the incomplete/partial program is shown in Figure~\ref{fig.pair_programming.example.input}. In the generated programs shown in Figures~\ref{fig.pair_programming.example.output_ChatGPT},~\ref{fig.pair_programming.example.output_GPT4},~and~\ref{fig.pair_programming.example.output_Tutor}, we have highlighted the lines where the input partial program has been updated. \TechChatGPT{} generates an incorrect program (i.e., the  attribute \emph{Correct} is $0$). \TechGPTFour{} generates a correct program and keeps the context (i.e., attributes \emph{Correct} and \emph{ContextKept} are $1$); however, it makes a large number of edits (\emph{EditLines} is $10$). \TechTutor{} generates a  correct program, keeps the context, and makes a small number of edits (i.e., \emph{EditLines} is $4$).

%

\begin{figure}[h!]
            \begin{subfigure}[b]{0.94\textwidth}
                \centering
                \scalebox{1}{
                \renewcommand{\arraystretch}{0.4}
                    \begin{tabular}{|p{1\linewidth}|}
                        \hline
                        \multicolumn{1}{|p{1\linewidth}|}{
\begin{lstlisting}[basicstyle=\fontsize{7.5}{7}\ttfamily,xleftmargin=1.65em,belowskip=-0.7em,aboveskip=-0em]
from math import sqrt
class Solution:
    def count_divisors(self, N):
        sq = int(sqrt(N))
        count = 0
        for i in range(1, sq + 1):
            if N % i == 0 and i % 3 == 0:
\end{lstlisting}}\\
                        \hline
                    \end{tabular}
                }
                \subcaption{Input: Partial program}
                \vspace{5mm}
                \label{fig.pair_programming.example.input}
            \end{subfigure}
            \begin{subfigure}[b]{0.94\textwidth}
                \centering
                \scalebox{1}{
                \renewcommand{\arraystretch}{0.4}
                    \begin{tabular}{|p{1\linewidth}|}
                        \hline
                        \multicolumn{1}{|p{1\linewidth}|}{
\begin{lstlisting}[basicstyle=\fontsize{7.5}{7}\ttfamily,xleftmargin=1.65em,belowskip=-0.7em,aboveskip=-0em,xleftmargin=1.65em,belowskip=-0.7em,aboveskip=-0.em,linebackgroundcolor={
                            \ifnum\value{lstnumber}=8\color{CodeHighlight}\fi
                            \ifnum\value{lstnumber}=9\color{CodeHighlight}\fi
                            \ifnum\value{lstnumber}=10\color{CodeHighlight}\fi
                            \ifnum\value{lstnumber}=11\color{CodeHighlight}\fi
                            }]
from math import sqrt
class Solution:
    def count_divisors(self, N):
        sq = int(sqrt(N))
        count = 0
        for i in range(1, sq + 1):
            if N % i == 0 and i % 3 == 0:
                count += 1
                if N // i != i and (N // i) % 3 == 0:
                    count += 1
        return count
\end{lstlisting}}\\
                        \hline
                    \end{tabular}
                }
                \subcaption{Output by \TechChatGPT}
                \vspace{5mm}
                \label{fig.pair_programming.example.output_ChatGPT}
            \end{subfigure}
            \begin{subfigure}[b]{0.94\textwidth}
                \centering
                \scalebox{1}{
                \renewcommand{\arraystretch}{0.4}
                    \begin{tabular}{|p{1\linewidth}|}
                        \hline
                        \multicolumn{1}{|p{1\linewidth}|}{
\begin{lstlisting}[basicstyle=\fontsize{7.5}{7}\ttfamily,xleftmargin=1.65em,belowskip=-0.7em,aboveskip=-0em,linebackgroundcolor={
                            \ifnum\value{lstnumber}=7\color{CodeHighlight}\fi
                            \ifnum\value{lstnumber}=8\color{CodeHighlight}\fi
                            \ifnum\value{lstnumber}=9\color{CodeHighlight}\fi
                            \ifnum\value{lstnumber}=10\color{CodeHighlight}\fi
                            \ifnum\value{lstnumber}=11\color{CodeHighlight}\fi
                            \ifnum\value{lstnumber}=12\color{CodeHighlight}\fi
                            \ifnum\value{lstnumber}=13\color{CodeHighlight}\fi
                            \ifnum\value{lstnumber}=14\color{CodeHighlight}\fi
                            \ifnum\value{lstnumber}=15\color{CodeHighlight}\fi
                            \ifnum\value{lstnumber}=16\color{CodeHighlight}\fi
                            \ifnum\value{lstnumber}=17\color{CodeHighlight}\fi
                            }]
from math import sqrt
class Solution:
    def count_divisors(self, N):
        sq = int(sqrt(N))
        count = 0
        for i in range(1, sq + 1):
            if N % i == 0:
                # If divisors are equal, increment count by one
                if N // i == i:
                    if i % 3 == 0:
                        count += 1
                else:  # Else increment count twice as we count both i and N/i
                    if i % 3 == 0:
                        count += 1
                    if N // i % 3 == 0:
                        count += 1
        return count
\end{lstlisting}}\\
                        \hline
                    \end{tabular}
                }
                \subcaption{Output by \TechGPTFour}		\vspace{5mm}
                \label{fig.pair_programming.example.output_GPT4}
            \end{subfigure}
            \begin{subfigure}[b]{0.94\textwidth}
                \centering
                \scalebox{1}{
                \renewcommand{\arraystretch}{0.4}
                    \begin{tabular}{|p{1\linewidth}|}
                        \hline
                        \multicolumn{1}{|p{1\linewidth}|}{
\begin{lstlisting}[basicstyle=\fontsize{7.5}{7}\ttfamily,xleftmargin=1.65em,belowskip=-0.7em,aboveskip=-0em,linebackgroundcolor={
                            \ifnum\value{lstnumber}=8\color{CodeHighlight}\fi
                            \ifnum\value{lstnumber}=9\color{CodeHighlight}\fi
                            \ifnum\value{lstnumber}=10\color{CodeHighlight}\fi
                            \ifnum\value{lstnumber}=11\color{CodeHighlight}\fi
                            }]
from math import sqrt
class Solution:
    def count_divisors(self, N):
        sq = int(sqrt(N))
        count = 0
        for i in range(1, sq + 1):
            if N % i == 0 and i % 3 == 0:
                count += 1
            if N % i == 0 and i != (N / i) and (N / i) % 3 == 0:
                count += 1
        return count
\end{lstlisting}}\\
                        \hline
                    \end{tabular}
                }
                \subcaption{Output by \TechTutor}		\vspace{5mm}
                \label{fig.pair_programming.example.output_Tutor}
            \end{subfigure}

    \caption{An illustrative example for the pair programming scenario on \PDivisors{} problem.}
	\label{fig.pair_programming.example}
\end{figure}
\clearpage

\subsection{Illustrative Example for Contextualized Explanation Scenario}
\label{app.sec.contextualizedexplanation}

Figure~\ref{fig.contextualized_explanation.example} shows an example for the contextualized explanation scenario on \PMergeStrings{} problem. In this example, the input program and a specific part of the program (line $11$) to be explained are shown in Figure~\ref{fig.contextualized_explanation.example.input}. Note that the input program has no bugs; see Footnote~\ref{footnote.contextualized}. In the generated explanations shown in Figures~\ref{fig.contextualized_explanation.example.output_ChatGPT},~\ref{fig.contextualized_explanation.example.output_GPT4},~and~\ref{fig.contextualized_explanation.example.output_Tutor}, we have highlighted the text where the explanation is wrong. Both \TechChatGPT{} and \TechGPTFour{} generate an incorrect explanation (i.e., the attribute \emph{Correct} is $0$). \TechTutor{}'s output meets all the quality criteria  (i.e., \emph{Overall} is $1$).

%

\begin{figure}[h!]
    \centering
    \begin{subfigure}[b]{0.5\textwidth}
        \centering
        \scalebox{0.85}{
            \renewcommand{\arraystretch}{1}
            \begin{tabular}{|p{1\linewidth}|}
                \hline
                \multicolumn{1}{|p{1\linewidth}|}{
\begin{lstlisting}[basicstyle=\fontsize{8.5}{7}\ttfamily,xleftmargin=1.65em,belowskip=-0.7em,aboveskip=-0.2em,linebackgroundcolor={
                            \ifnum\value{lstnumber}=11\color{CodeHighlight}\fi}
                    ]
class Solution:
    def merge(self, S1, S2):
        x = ''
        z = S1 + S2[::-1]
        i = 0
        j = len(z) - 1
        n = min([len(S1), len(S2)])
        for k in range(0, n):
            if i != j:
                x += z[i]
                x += z[j]
            
            i += 1
            j -= 1
        if len(S1) > len(S2):
            x = x + S1[n:]
        else:
            x = x + S2[n:]
        return x
\end{lstlisting}}\\
                \hline
            \end{tabular}
        }
        \subcaption{Input: Program and a specific part (line $11$)}				
        \vspace{5mm}
        \label{fig.contextualized_explanation.example.input}
    \end{subfigure}
    \allowbreak
    
    \begin{subfigure}[t!]{0.9\textwidth}
    {
        \centering
        \scalebox{0.9}{
            \renewcommand{\arraystretch}{1.6}
            \begin{tabular}{|p{1\linewidth}|}
                \hline
                \multicolumn{1}{|p{1\linewidth}|}{
                    \input{figs/7_contextualized_explanation/fig_examples_data_new/Q11_p4_ChatGPT}
                    \vspace{1mm}
                }\\
                \hline
            \end{tabular}
        }
        \caption{Output by \TechChatGPT}
        \vspace{5mm}
        \label{fig.contextualized_explanation.example.output_ChatGPT}
    }
    \end{subfigure}
    
    \begin{subfigure}[t!]{0.9\textwidth}
    {
        \centering
        \scalebox{0.9}{
            \renewcommand{\arraystretch}{1.6}
            \begin{tabular}{|p{1\linewidth}|}
                \hline
                \multicolumn{1}{|p{1\linewidth}|}{
                    \looseness-1\input{figs/7_contextualized_explanation/fig_examples_data_new/Q11_p4_GPT4}
                    \vspace{1mm}
                }\\
                \hline
            \end{tabular}
        }
        \caption{Output by by \TechGPTFour}
        \vspace{5mm}
        \label{fig.contextualized_explanation.example.output_GPT4}
    }
    \end{subfigure}
    
    \begin{subfigure}[t!]{0.9\textwidth}
    {
        \centering
        \scalebox{0.9}{
            \renewcommand{\arraystretch}{1.6}
            \begin{tabular}{|p{1\linewidth}|}
                \hline
                \multicolumn{1}{|p{1\linewidth}|}{
                     \vspace{-4mm}
                     \input{figs/7_contextualized_explanation/fig_examples_data_new/Q11_p4_Tutor}
                     \vspace{1mm}
                }\\
                \hline
            \end{tabular}
        }
        \caption{Output by \TechTutor}
        \vspace{5mm}
        \label{fig.contextualized_explanation.example.output_Tutor}
    }
    \end{subfigure}

    \caption{An illustrative example for the contextual explanation scenario on \PMergeStrings{} problem.}
	\label{fig.contextualized_explanation.example}
\end{figure}

\clearpage

\subsection{Illustrative Example for Task Synthesis Scenario}
\label{app.sec.taskcreation}

\looseness-1Figure~\ref{fig.task_creation.example} shows an example for the task synthesis scenario on \PGCD{} problem. In this example, the student's buggy program shown in Figure~\ref{fig.task_creation.example.input} has a mistake in passing the arguments to the recursive function call. \TechChatGPT{}'s output is of bad quality w.r.t. the attributes \emph{SimilarBugs} and \emph{MinimalBugs}. \TechGPTFour{}'s output is of bad quality w.r.t. the attribute \emph{SimilarBugs} as there is no bug in the generated program as required. Interestingly, \TechGPTFour{} also writes a detailed textual description---not shown here---where it \emph{wrongly reasons about} how it has generated new problem and new buggy program that matches the prompt requirements. \TechTutor{}'s output meets all the quality criteria  (i.e., \emph{Overall} is $1$). 

%

%

\begin{figure}[h!]
    \centering
    \begin{subfigure}[b]{0.6\textwidth}
    {
        \centering
        \scalebox{0.85}{
            \renewcommand{\arraystretch}{1}
            \begin{tabular}{|p{1\linewidth}|}
                \hline
                \multicolumn{1}{|p{1\linewidth}|}{
\begin{lstlisting}[basicstyle=\fontsize{8.5}{5}\ttfamily,xleftmargin=1.65em,belowskip=-1em,aboveskip=-0.2em]
class Solution:
    def gcd(self, A, B):
        if B == 0:
            return A
        else:
            return self.gcd(A, B % A)
\end{lstlisting}}\\
                \hline
            \end{tabular}
        }
        \vspace{-1mm}
        \subcaption{Input: Buggy program}	
        \vspace{4mm}
        \label{fig.task_creation.example.input}
    }
    \end{subfigure}      
    \allowbreak
    
    \begin{subfigure}[t!]{\textwidth}
    {
        \centering
        \scalebox{0.85}{
            \renewcommand{\arraystretch}{1.6}
            \begin{tabular}{|p{1\linewidth}|}
    \hline
    \multicolumn{1}{|c|}{\componentheader{New Problem}
    \vspace{-1mm}}\\
    \multicolumn{1}{|p{1\linewidth}|}{
        Sum of Two Numbers: Given two positive integers A and B, find the sum of A and B. You don't need to read input or print anything. Complete the function `sum\_of\_two\_numbers()' which takes two positive integers as input parameters and returns an integer.\vspace{-2.5mm}\newline
        \newline
        Example: Input: A = 4, B = 5. Output: 9. Explanation: The sum of 4 and 5 is 9.\newline
        \vspace{-2mm}
    }\\
    \hline
    \multicolumn{1}{|c|}{\componentheader{New Buggy Program}}\\
    \multicolumn{1}{|p{1\linewidth}|}{ 
        \vspace{-1mm}
\begin{lstlisting}[basicstyle=\fontsize{8.5}{5}\ttfamily,xleftmargin=1.65em,belowskip=-1em,aboveskip=-0.9em]
class Solution:
    def sum_of_two_numbers(self, A, B):
        if A == 0:
            return B
        else:
            return self.sum_of_two_numbers(B, A - 1)
\end{lstlisting}
    }\\
    \hline
\end{tabular}
        }
        \vspace{-1mm}
        \caption{Output by \TechChatGPT}
        \vspace{4mm}
        \label{fig.task_creation.example.output_ChatGPT}
    }
    \end{subfigure}
    
    \begin{subfigure}[t!]{\textwidth}
    {
        \centering
        \scalebox{0.85}{
            \renewcommand{\arraystretch}{1.6}

\begin{tabular}{|p{1\linewidth}|}
    \hline
    \multicolumn{1}{|c|}{\componentheader{New Problem}
    \vspace{-1mm}}\\
    \multicolumn{1}{|p{1\linewidth}|}{
        Multiplication of Two Positive Numbers: Given two positive integers A and B, return the result of A multiplied by B. You don't need to read input or print anything. Complete the function multiply() which takes two positive integers as input parameters and returns an integer.\vspace{-2.5mm}\newline
        \newline
        Example 1: Input: A = 3, B = 6. Output: 18. Explanation: 3 multiplied by 6 is 18.\newline
        Example 2: Input: A = 2, B = 5. Output: 10. Explanation: 2 multiplied by 5 is 10.\newline
        \vspace{-2mm}
    }\\
    \hline
    \multicolumn{1}{|c|}{\componentheader{New Buggy Program}}\\
    \multicolumn{1}{|p{1\linewidth}|}{ 
        \vspace{-1mm}
\begin{lstlisting}[basicstyle=\fontsize{8.5}{5}\ttfamily,xleftmargin=1.65em,belowskip=-1em,aboveskip=-0.9em]
class Solution:
    def multiply(self, A, B):
        if B == 1:
            return A
        else:
            return A + self.multiply(A, B-1)
\end{lstlisting}
    }\\
    \hline
\end{tabular}

        }
        \vspace{-1mm}
        \caption{Output by \TechGPTFour}	
        \vspace{4mm}
        \label{fig.task_creation.example.output_GPT4}
    }
    \end{subfigure}
    
    \begin{subfigure}[t!]{\textwidth}
    {
        \centering
        \scalebox{0.85}{
            \renewcommand{\arraystretch}{1.6}
            \begin{tabular}{|p{1\linewidth}|}
    \hline
    \multicolumn{1}{|c|}{\componentheader{New Problem} \vspace{-1mm}}\\
    \multicolumn{1}{|p{1\linewidth}|}{        
        Euclidean Division: Find the reminder of A divided by B, using successive subtractions. A and B are positive integers.
        \vspace{-2.5mm}\newline
        \newline
        Example 1: Input: A = 5, B = 3. Output: 2\newline
        Example 2: Input: A = 10, B = 3. Output: 1\newline
        Example 3: Input: A = 10, B = 5. Output: 0\newline
        \vspace{-2mm}
    }\\
    \hline
    \multicolumn{1}{|c|}{\componentheader{New Buggy Program}}\\
    \multicolumn{1}{|p{1\linewidth}|}{ 
        \vspace{-1mm}
\begin{lstlisting}[basicstyle=\fontsize{8.5}{5}\ttfamily,xleftmargin=1.65em,belowskip=-1em,aboveskip=-0.9em]
class Solution:
    def reminder(self, A, B):
        if A < B:
            return A
        return self.reminder(B, A-B)
\end{lstlisting}
    }\\
    \hline
\end{tabular}

        }
        \vspace{-1mm}
        \caption{Output by \TechTutor}	
        \label{fig.task_creation.example.output_Tutor}
    }
    \end{subfigure}

        \caption{An illustrative example for the task synthesis scenario on \PGCD{} problem.}
	\label{fig.task_creation.example}
\end{figure}
\clearpage


\end{appendices}

\end{document}